\documentclass[12pt]{article}
\usepackage{epsfig}

\newcommand{\beq}{\begin{eqnarray}}
\newcommand{\eeq}{\end{eqnarray}}

\begin{document}

\begin{flushright}
TPJU-15/97 \\
December 1997
\end{flushright}

\begin{center}
\vspace{24pt}

{\Large \bf Lattice Models of Random Geometries
\footnote{This paper is presented as a qualifying
thesis for the habilitation at the Jagellonian
University}}
\vspace{24pt}

{\large \sl Zdzis\l{}aw Burda \footnote{email~: burda@thp1.if.uj.edu.pl}}
\vspace{10pt}

Institute of Physics, Jagellonian University, \\
ul. Reymonta 4, PL 30-059 Krakow, Poland
\vspace{10pt}

\begin{abstract}
We review models of random geometries based on the 
dynamical lattice approach. We discuss 
one dimensional model of simplicial complexes 
(branched polymers), two dimensional model of 
dynamical triangulations and four dimensional model
of simplicial gravity. 

\medskip
\noindent
PACS~: 04.60.Nc; 05.70.Fh
\end{abstract}

\end{center}

\newpage

\section*{Introduction}
\setcounter{equation}{0}

The theory of random geometries provides a 
useful framework to describe a wide
spectrum of problems in statistical physics ranging 
from physics of polymers, membranes and
domain walls, to problems where the dynamical 
geometry arises as a purely mathematical object. 
Another area of applications of the theory 
is related to the  quantization of the geometrical theories 
like the string theory or the general relativity 
\cite{pb,p1,w6,h,w,am1,aj1}. 

Nowadays, one can not imagine physics
without the quantum theory or the general relativity.
Both theories describe physical phenomena with 
a remarkable accuracy and both have proven to have
great predictive power. Each of them has its own
domain of applicability and so far there is no
experiment which would contradict either of them.
Both theories do not interfere with each other.
Theoretical consistency requires, however, that there
exists a covering theory which would unify the quantum
theory and general relativity. This means that either 
gravity must be quantized or quantum theory "gravitized". 
The formulation of such a theory is one of the greatest challenges 
of theoretical physics. It attracts researchers from different
areas of physics and results in numerous independent 
approaches \cite{gph}. 

The theory of random geometries reported here is
a generalization of the conventional Euclidean path 
formalism successfully used as a nonperturbative
method in field theory. The great interest in the theory
of random geometries in the last decade was triggered
by string theory (see for review \cite{d1,a1,gm2}).
The representation of the quantum amplitudes for
strings in terms of the amplitudes for 2d gravity 
coupled minimally to matter fields
evolved, in parallel to the string interpretation,
as a theory of 2d quantum gravity \cite{pb,p1}. 
The success of the two dimensional theory \cite{kpz,d2,dk}
and especially of the dynamical triangulation 
approach \cite{d3,kkm,adf}
which, in particular, allowed for addressing 
nonperturbative questions \cite{bk1,ds,gm}, 
and for calculating invariant correlations
\cite{kkmw,aw}, challenged researchers to generalize 
the ideas to higher dimensional gravity.
The generalization was done stepwise~: first to three
dimensions \cite{am2,bk2,av,abjk1}, then to four 
\cite{am1,aj1,abjk2,ckr1,dbs,bm}. 

In this review paper we focus on the discretized 
approach combining the lattice regularization
with the standard concepts of critical phenomena
in statistical physics. 
The paper is organized as follows. After a short
introduction where we recall the 
basic concepts and the discretization scheme,
in the successive sections we discuss the statistical physics
of one dimensional simplicial complexes (branched polymers),  
the
theory of random surfaces and four dimensional simplicial
gravity. 

The model of branched polymers is solvable \cite{adfo,adj1,b1,bb1,jk}.
It undergoes a phase transition \cite{bb1}
related to the collapse of geometry and to the appearance
of singular vertices \cite{bbj}. An analogous phase transition 
is also encountered in models of random surfaces \cite{adfo,k1}
and higher dimensional simplicial gravity \cite{hin,ckrt,bbpt}. 
The mechanism of the transition can be mapped  onto
the condensation of the balls-in-boxes 
model \cite{bbj,bb2,bbj2}. 

The model of dynamically triangulated surfaces 
is also well understood.  It has three phases~: collapsed geometries, 
branched polymers and 2d Liouville gravity \cite{k1,bkkm,adf2}. 
The gravity phase has a well defined continuum 
limit corresponding to the quantum Liouville field theory.
The theory is analytically solvable by means of the continuum formalism 
\cite{kpz,d2,dk} and by the discretized approach using matrix 
model techniques \cite{kkm,k2,d4}. The scaling and universal 
properties of this phase are well established. 

Our understanding of four dimensional simplicial gravity
has improved recently. We have gained insight into
the phase structure of the model and
the mechanisms governing the behaviour of the system. 
Nevertheless, we are still far from achieving the ultimate 
goal of the study, namely, the determination of the relation of 
simplicial gravity to the continuum physics. 

The basic difficulty encountered in investigating 
the model is the lack of analytic methods of summing over 
four dimensional geometries.  For the time being the only
way of studying the model is the Monte Carlo technique 
\cite{bbj3,db2}. 

By means of this method, the basic properties of the model 
have been determined.  We discuss the state 
of art and summarize the main properties of the model in
the section on four dimensional gravity. 
We show that the model possesses 
a well defined thermodynamic limit \cite{aj3,bm2,bbp} 
We discuss the phase structure. The model has
two phases~: the collapsed phase with infinite Hausdorff 
dimension and the elongated phase with the Hausdorff 
dimension equal two. It was believed that 
the phase transition  between
the elongated and the crumpled phase
under a variation of the gravitational coupling 
constant was second order.
Massive numerical simulations showed that the transition is
however discontinuous, meaning that one can not associate 
a continuum physics with the critical point 
\cite{bbkp,db1}. 
The discontinuity of the transition may be explained 
in terms of the constrained mean field scenario \cite{bb2}. 
A physical explanation advocated in \cite{jk,amm}
is that the conformal mode gets released at the transition due to
the entropical dominance of spiky configurations, similarly as above 
the $c=1$ barrier in two dimensions \cite{c,k3,d5}. According to this, 
if one extends the model by adding matter fields, 
there may exist another phase, like the Liouville phase 
in two dimensions, free of this instability. 
Indeed, recent numerical investigations 
of 4d simplicial gravity interacting with 
vector fields support this scenario \cite{bbkptt}.

Apart from the main line of presentation
we discuss the balls-in-boxes model which serves 
as a mean field approximation for the dynamical lattice 
models \cite{bbj,bbpt}. We also sketch ideas underlying 
the Monte Carlo simulations of dynamical lattices. 
We end the paper with a short summary.

\section{Preliminaries}
\setcounter{equation}{0}

One defines the partition function on the
ensemble of geometries $\{ G\}$~: 
\begin{equation}
Z = \sum_{G} W[G] \, ,
\label{dez}
\end{equation}
where $W[G]$ is a nonnegative weight function
given by the Gibbs measure. In the quantization
procedure of geometrical theories, the weight
is  
\beq
W[G] = e^{-S[G]}
\eeq
where $S[G]$ is the Euclidean action. In this case,
the statistical sum (\ref{dez}) corresponds to the quantum
amplitudes. The simplest example of a model belonging
to this class is a model of random paths. 
The sum over random paths gives the free particle 
propagator as one expects from the quantum theory \cite{pb}.

In general, the construction of the theory follows the ideas
of statistical field theory. There are however many detail
differences.  A standard field theory is defined on a
manifold with an inert geometry. There is no such 
fixed underlying structure which would serve as 
a reference manifold in the models of random geometry.  
Geometrical quantities like a distance or a Hausdorff 
dimension are the dynamical properties of a given 
ensemble, not of a single manifold as in field theory.
In field theory one defines field correlators
of the type $\langle\phi(x)\phi(y)\rangle$ where $x,y$ 
are the points on the basis manifold. One investigates then the
behaviour in terms of the distance $|x-y|$. From this behaviour one 
can learn about the excitations in the system. 
The task is more difficult in the case of the random geometries 
where one cannot fix the points $x,y$ because manifolds fluctuate.
It is even hard to define the correlators between the pairs of
points at a given distance $r$, since the distance between
points is a global property of the manifold. 
The correlation functions can be found analytically only in
few particular cases like branched polymers \cite{adj1,b1} or
two dimensional pure gravity. In the latter case the 
calculation requires an elaborate technique which was 
developed only for this purpose \cite{kkmw,aw}.

There are many new interesting phenomena not present in
field theory, like the geometrical collapse, 
the back reaction of matter fields on geometry
or the change of the dimensionality of the system.

The class of geometries which can be considered in 
the partition function (\ref{dez}) is not limited to 
paths, surfaces or higher dimensional hyper--surfaces.
Geometry can be thought of as any set with a given
metric structure such as a diagram 
with a function specifying the distance between vertices. 

A theory is said to be geometrical if the
action $S[G]$ is a function of geometry only. In calculations,
one usually uses redundant representations
of geometry. Redundancy manifests itself as a
gauge symmetry of the action.
For random paths, for example,
a path between two points $a,b$ 
can be represented as a continuous map of the unit 
interval onto a target space where the path is embedded~: 
$t \rightarrow x[t]$, $a=x[0]$, $b=x[1]$. The path will not change if one 
changes the map to~: $t \rightarrow y[t] = x[f(t)]$ 
where $f$ is a monotonic diffeomorphism preserving 
the ends of the unit interval $f(0)=0$, $f(1)=1$. 
A geometrical action will not change either~: 
$S[x[t]] = S[y[t]]$.  
This diffeomorphism invariance corresponds to the gauge symmetry
of the representation.

In general for any representation, a set of maps which
represent the same geometry $G$, defines an equivalence class
(gauge orbit).  If one represents geometry by 
a metric tensor $g_{\mu\nu}$, the geometrical action must be 
invariant with respect to diffeomorphism. 
The simplest invariants yield the Einstein-Hilbert action~:
\beq
S[g_{\mu\nu}] = \lambda \int {\rm d}^d \xi \sqrt{g} - 
\frac{1}{2\pi G} \int {\rm d}^d \xi \sqrt{g} R
\label{deeh}
\eeq
where $R$ is the scalar curvature. 
The symmetry of this representation divides 
the metric tensors into diffeomorphism classes.
The sum (\ref{dez}) must be defined in such a way
that the over-counting of metrics from the same
diffeomorphism class is avoided. In general there are two 
strategies to do this. Either one picks up 
only one representative of each equivalence class, 
which is usually technically impossible,
or one sums over all elements of each equivalence 
class but at the same time one divides
out the volume of the equivalence class 
(gauge orbit)~:
\beq
Z = \sum_{G} W[G] = \sum_T \frac{1}{C[T]} W[T]
\label{der}
\eeq
A practical realization of this idea was elaborated
in field theory as the gauge fixing procedure. 
The role of the factor $1/C[T]$ is played by 
the Fadeev-Popov determinant. The gauge fixing 
procedure was successfully carried out for random 
paths and random surfaces corresponding to 2d gravity interacting 
minimally with conformal matter with $c\le 1$ \cite{pb,p1,kpz,d2,dk}.

For continuous geometries, the expression 
(\ref{der}) has  only a symbolic meaning. 
One has to express the sums in (\ref{der})
in terms of well defined mathematical quantities
which would uniquely specify what is really meant by 
the sums.
To this purpose one introduces a short 
distance cut-off.  In the continuum 
approach, one develops the theory covariantly and
in the end one introduces the cut-off  to compute  
the final covariant integrals.
On the contrary, in the lattice regularization,
one introduces the cut-off at the very beginning, 
before starting the whole machinery. In this way one
breaks symmetries of the continuum theory but one hopes
to recover them in the continuum limit when the cut-off 
is carefully sent to zero. The lattice, which is an 
auxiliary construct, is eventually removed from 
the problem in this limit.

The idea of a discretization is commonly used in
the statistical field theory.
It is also not new in the context of geometrical
models where it is known under the name Regge 
calculus \cite{r}. 
The Regge's idea was to use a piecewise linear manifold
to approximate the continuous geometry. Regge's lattices have a fixed
connectivity. Geometrical degrees of freedom are encoded
in the link lengths.
This idea has proven to be very helpful on the classical level 
where it is just the finite element method 
to solve the classical field equations. 
Its applicability to define the sum (\ref{dez}) over 
random geometries is, however, limited by 
the integration measure problem. In the two 
dimensional case, the integration measure
can be deduced from the Liouville theory. It turns
out that it is given by a highly nonlocal expression 
involving all link variables of the triangulation \cite{m}. 
In general, the measure is not known for Regge 
manifolds. Any attempt to mimic the measure 
by local expressions fails \cite{hj}. In other words the Regge method 
is well suited as a method to approximate Riemannian structures 
but not to approximate the sum over them.

An alternative discretized approach to random geometries
is based on dynamical triangulations \cite{d3,kkm,adf}.
The randomness of the geometry is encoded in the fluctuating 
connectivity of the lattice. Link lengths are fixed.
The dynamical triangulations method is not as good
as the Regge calculus as an approximation of
the classical equations but it is very well suited 
to problem of summing over random geometries. 

The basic Ansatz is that the sum over
diffeomorphism classes of the continuum approach
can be regularized as a sum over equilateral 
triangulations. The volume of the symmetry group 
for such a triangulation with $N$ labelled vertices 
is equal to the number of possible relabellings 
$C[{\cal T}]=N!$~:
\beq
\int \frac{{\cal D} g_{\mu\nu}}{{\cal D} {\rm diff}} \dots \,
\mapsto \, \sum_{{\cal T}} \frac{1}{C[{\cal T}]} \dots \, =  \,
\sum_{\cal T} \frac{1}{N!} \dots \, .
\label{dem} 
\eeq
If one uses the non labelled triangulations $T$ instead,
the symmetry factor $C[T]$ is equal to $N!$ divided
by the number of distinct labelling of the triangulation. 
For triangulations without any symmetry, the factor $C[T]$
is equal one.  

The sum over dynamical triangulations can be 
done analytically by the matrix model technique.
The results of the matrix model and the Liouville 
theory agree. This is usually treated as a strong 
indication that the two methods
provide correct definitions of the integration measure
over random surfaces. An advantage of the lattice method 
is that without changing the Ansatz (\ref{dem})
one can generalize it beyond the Liouville phase 
of two dimensional gravity, in particular, 
to higher dimensional gravity.

If one applies the Regge calculus to the discretization
of the Einstein-Hilbert action (\ref{deeh}) 
on the equilateral simplicial lattice one obtains \cite{am1,aj1}~:   
\beq
S = {\kappa_d} N_d - {\kappa}_{d-2} N_{d-2}
\label{deda}
\eeq
where $N_d$ and $N_{d-2}$ denote the number of $d$-simplices
and $(d-2)$-simplices of the simplicial manifold. 
The coupling constants $\kappa$'s are related
to the couplings of the continuum action (\ref{deeh})
and to the lattice spacing in the naive continuum limit.

One can extend this discretization procedure to 
geometrical actions with higher derivative terms \cite{ajk,abjk3}
or actions which describe interaction of gravity
with some matter fields \cite{abjk1,abjk2,ckr3,ckr4}. 
This discretization scheme 
leaves some freedom because one can add  
to the discretized action terms, which disappear  
in the naive continuum limit. One should perhaps look at the
discrete models from a different perspective, in which
the model is not viewed as a discretization of a continuum theory
but rather as a primary definition of the theory. Then if one finds 
a continuum limit then one can ask what is the underlying continuum theory 
related to this limit, and whether this theory is related to gravity.
In fact this is the most difficult part of the procedure, 
since we do not have the experimental data.
All we know about the continuum theory is that it should
reproduce general relativity in the classical limit.
This, together with the self consistency requirement, 
provides the only available checks for the consistency
of the constructed theory.

\section{Branched polymers}
\setcounter{equation}{0}

The model of branched polymers provides a useful ground for
testing various ideas. It is simply and solvable. It plays 
a similar role for random geometries as the Ising model 
for statistical field theory \cite{adfo,adj1,b1,bb1,jk}. 
The model undergoes a phase
transition related to the change of the Hausdorff
dimension and to the collapse of geometry \cite{bb1,jk,bbj}.
A similar transition is also
present in the models of random surfaces \cite{k1}
and four dimensional gravity \cite{hin,ckrt}.

Branched polymer is a graph (one dimensional simplicial
complex) without loops. Geometry on a branched polymer
is given by the geodesic distance between vertices of the graph.
The distance is defined as the number of links 
of the shortest path between the points.  On a tree-like 
graph there is only one path joining any two points.

In the simplest case, one considers polymers whose
vertices are independent in the sense that there is
no direct correlation between the branching orders.
A branching order is defined as the number of links 
emerging from the vertex. 
The action for such a branched polymers is
\beq
S[T] = \mu N[T] - \sum_{i\in T} s(q_i)
\label{bpa}
\eeq
where $N[T]$ is the number of vertices on a branched polymer $T$,
$\mu$ is the chemical potential and $s(q)$ is a one-vertex
action depending only on the vertex order $q$.
The sum in the second term of (\ref{bpa})
runs over all vertices of the branched polymer.
The grand canonical partition function defined
on the ensemble of trees of arbitrary size, reads~:  
\beq 
Z(\mu) = \sum_{T} \frac{1}{C[T]} 
\exp \Big( \sum_{i\in T} s(q_i) - \mu N[T] \Big) = \sum_N z(N) e^{-\mu N} \, .
\label{bpz}
\eeq
and can be treated as a discrete Laplace transform
of the canonical partition function $z(N)$,
being the statistical sum over the ensemble of trees
with $N$ vertices.  

It can be shown that the coefficients 
$z(N)$ grow exponentially 
\beq
z(N) \sim N^{\gamma-3} \exp (\mu_{cr} N) \, ,
\label{bpzn}
\eeq
for large $N$. This means that the thermodynamic limit,
$N\rightarrow \infty$, is well defined. The quantity
$\mu_{cr}$ is the critical value of the chemical potential 
at which the grand-canonical function has a singularity. 
The power-like corrections $N^{\gamma-3}$
determine the type of the singularity of the 
grand canonical partition function 
for $\Delta \mu = \mu - \mu_{cr} \rightarrow 0^+$~:
$Z(\mu) \sim \Delta \mu^{2-\gamma}$. 
The singular part of the susceptibility 
defined as the second derivative of $Z$,
behaves as 
\beq
\chi(\mu) = Z''(\mu) \sim \Delta \mu^{-\gamma}
\label{chi1}
\eeq
at the critical $\mu$. The exponent $\gamma$
is called a susceptibility exponent or entropy exponent.
The exponent $\gamma$ is generally used to determine 
the universality class of models of random geometries \cite{kpz,d2,dk}. 

The susceptibility is related to the 
puncture-puncture correlation function defined
as \cite{aw,adj1}~:
\beq
G(r,\mu) = \sum_{T} \frac{1}{C[T]} {\rm e}^{-S[T]}
\sum_{a,b} \delta(r - d(a,b))  
\label{bpg}
\eeq
The internal sum runs over all pairs of points.
The delta function selects contributions  from
pairs at a distance $r$. Integrated over $r$, the
two-point correlator gives~:
\beq
\chi(\mu) = \sum_r G(r,\mu)
\label{chi2}
\eeq
For large $r$, the correlation function $G(r,\mu)$ 
falls off exponentially and this is a general feature
of models of random geometries \cite{ajw}.
One associates a mass with the exponential fall-off~:
\beq
m = - \lim_{r\rightarrow\infty} \ln \frac{G(r,\mu)}{r} \, .
\eeq
This mass is directly related to the large $N$ limit. 
It is physically important if the mass scales to 
zero when $\Delta \mu = \mu - \mu_{cr} \rightarrow 0^{+}$.
If it does, the geometry has a well defined Hausdorff dimension.
The dimension is related to the mass critical index given
by the scaling formula \cite{ajw}~:
\beq
m \sim \Delta \mu^{1/d_H} .
\label{mass}
\eeq
If it does not scale, the geometry is collapsed, 
as we show later. 
To see that the exponent $d_H$ may be indeed identified 
with the Hausdorff dimension in the scaling case,
it is convenient to consider a counterpart of
the puncture-puncture correlator (\ref{bpg}) 
in the canonical ensemble with a fixed size $N$. Similarly
to the partition functions (\ref{bpz}) one can
relate the correlation functions by the
discrete Laplace transform~:
\beq
G(r,\mu) = \sum_N G(r,N) e^{-\mu N} 
\label{bpcc}
\eeq
where $G(r,N)$ is the correlation function in the
canonical ensemble. Defined this way, $G(r,N)$ 
has an unnatural normalization proportional to $z(N)$.
One can get rid of it, by defining the normalized correlator~:
\beq
g(r,N) = \frac{G(r,N)}{G(0,N)} 
= \frac{1}{N} \big\langle \sum_{a,b} \delta(r - d(a,b)) \big\rangle_N  
\eeq
where the averaging $\langle \dots \rangle_N$ is over the ensemble 
of trees of size $N$. 
Now we are interested in the large $N$ behaviour of the function $g$.
This behaviour can be extracted from the inverse Laplace transform
of $G(r,\mu) \sim \exp ( -\Delta \mu^{1/d_H} r) $,
(\ref{bpg}) for $\Delta \mu \rightarrow 0$.
Small $\Delta \mu$ corresponds to large $N \sim 1/\Delta \mu$ in
the transform, whence the normalized correlation function $g(r,N)$
must be a function of the argument $r/N^{1/d_H}$ for large $N$.
The function $g(r,N)$ measures the average
number of points at a distance $r$ from a random point.
Summed over $r$, $g(r,N)$ gives the number of all points $N$.
If one inserts the universal argument $r/N^{1/d_H}$
one obtains, that the average distance between points~:
\beq
\langle r \rangle = \frac{1}{N} \sum_r  r \, g(r,N) 
\label{bpxi}
\eeq
behaves for large $N$ as
\beq 
\langle r \rangle \sim N^{1/d_H} \, .
\label{bpdH}
\eeq
The last formula relates the size of the system $N$ to 
the typical linear extension of the system, naturally
leading to the interpretation of the
mass exponent $d_H$ (\ref{mass}) 
as the Hausdorff dimension.

Since the model is solvable, the critical indices~:
the susceptibility exponent $\gamma$ and 
the Hausdorff dimension $d_H$ can be calculated.

In practical calculations one considers 
the ensemble of planar rooted trees \cite{bb1}. 
Rooted trees have one marked vertex with one attached link. 
The planarity 
means that trees are drawn on a plane. The existence 
of the root and the planarity uniquely specify the 
symmetry factor $C[T]$. One can find a representation 
where $C[T]=1$.

The grand-canonical partition function (\ref{bpz})  
is given by the solution of the equation \cite{adfo,bb1}~:
\beq
\mu = K(Z) =  \log \sum_{q=1}^\infty p(q) Z^{q-2}
\label{series}
\eeq
which is to be solved for $Z$. The series coefficients
$p(q)$ are related to the one-vertex action~: 
$p(q) = {\rm e}^{-s(q)}$. A vertex with order $q$
contributes $p(q)$ to the total weight of the
tree. The weights $p$'s are nonnegative in unitary models.
Vertices with the order $q$ are forbidden if $p(q)=0$. 
One requires $p(1)>0$ 
in order to have the endpoints in the tree. 
In order that the polymer could branch,
at least one weight must be positive for
$q\ge 3$.

The solution of the equation (\ref{series}) for $Z$ is
given by the inverse function of $K$~: $Z = K^{-1}(\mu)$. 
In fact, without inverting one can find  the
singularities of $Z$. Namely $Z$ is singular at a certain 
$\mu_{cr}$ when either the derivative of the inverse function
is zero, $K'=0$, or the inverse function $K$ is itself singular 
at $Z_{cr} = Z(\mu_{cr})$ \cite{bb1}. 

In the former case one gets~:
\beq
\mu = K(Z_{cr}) + \frac{1}{2} K''(Z_{cr}) (Z - Z_{cr})^2 + \dots
\eeq
which gives $Z \sim \Delta \mu^{1/2}$, where $\Delta \mu = \mu - K(Z_{cr})$. 
It follows that $\gamma=1/2$,
as seen from the comparison with 
the singularity $Z \sim \Delta \mu^{1-\gamma}$ 
of the partition function\footnote{In a rooted ensemble, like
in the one considered here, one vertex
is fixed by the root. Therefore to obtain susceptibility $\chi$
one needs to differentiate $Z$ with respect to $\mu$ 
only once and not twice as in (\ref{chi1}) where $Z$  is
partition function for a non-rooted ensemble.}.
This is a generic situation since
the function $K$ has a minimum  
for any choice of weights $p$'s fulfilling
the general assumptions discussed before. 

The function $K$ is singular when the series (\ref{series}) 
has a finite radius of convergence. For example, 
for the weights, which for large $q$ behave as 
$p(q) \sim q^{-\beta}$, the series has a
singularity $(1-Z)^{\beta-1}$ at $Z=1$. When $\beta > 2$,
the situation is interesting since there are three different possible  
behaviours of the singular part of the grand canonical
partition function. 
Let $Z_0$ be the value at which the derivative of $K$ vanishes~:
$K'(Z_0)=0$. Now we have two special points $Z=Z_0$ and
$Z=1$. When  $Z_0<1$, the singularity of $Z$ comes from inverting 
$K$ around $Z_0$. In this case one obtains the generic value $\gamma=1/2$,
as previously. When $Z_0>1$ the singularity of $Z$ comes from the
singularity of $K$ at $Z = 1$. In this case 
$Z \sim \Delta \mu^{\beta-1}$ and hence $\gamma = 2-\beta$.
This is a semi-generic situation. 
Finally, in the marginal situation when $Z_0=1$, the singularity of the 
partition function is $Z \sim \Delta \mu^{1/(\beta-1)}$ 
when $2<\beta<3$ or $Z \sim \Delta \mu^{1/2}$ otherwise. Hence
the exponent $\gamma$ is equal to $(\beta-2)/(\beta-1)$ or  $1/2$ 
respectively.

It is worth noting that the situation will not change if
we add an exponential prefactor to the weights $p(q)$~:
$q^{-\beta} \rightarrow e^{-\kappa q} q^{-\beta}$. 
Namely, due to the Euler relation, the sum of the 
vertex (branching) orders over the non-rooted vertices is~:
\beq
\sum_{i} q_i = 2N-1 \, ,
\eeq
and therefore the exponential term can be absorbed
in the cosmological term ${\rm e}^{-\mu N}$ 
(\ref{bpz}) by a redefinition of the chemical potential
$\mu \rightarrow \mu + 2\kappa$. After the
redefinition the weights have again the original form
$p(q) \sim q^{-\beta}$ and we obtain the previous
situation.

The generic phase of branched polymers has many universal
properties. In particular the normalized 
puncture-puncture 
correlation function has for large $N$ the form~:
\beq
g(r,N) = \sqrt{N} a \, \bar{g}\big(\frac{ar}{\sqrt{N}}\big)
\label{bpg1}
\eeq
where $\bar{g}$ is the universal function given by the formula~:
\beq
\bar{g}(x) = 2 x e^{-x^2}
\label{bpg2}
\eeq
of the universal argument $x = a r/N^{1/2}$. 
The form of the function $\bar{g}$ does not depend on the 
choice of $p$'s as long as the system is in the generic phase. 
The universality tells us that the local properties like the
branching distribution do not affect the long range behaviour.
The change of the weights $p$'s can be compensated by the change
of only one parameter $a$ in the correlation function. 
 
The Hausdorff dimension is $d_H=2$, as follows from the
universal scaling $x = a r /N^{1/2}$. Finite size
calculations show that the scaling is weakly broken
for finite $N$ and one should use a shifted argument
$x = a(r + \delta)/N^{1/2}$ \cite{b1,aj2}.   

In the marginal situation the form of the correlation function
changes and so does the universal parameter \cite{jk}. 
In this case the Hausdorff dimension is $d_H = 1/\gamma$ which is \,
$d_H=(\beta-1)/(\beta-2)$ for $2<\beta\le3$ \, or \, $d_H=2$ \, otherwise.  
In the semi-generic situation the normalized correlation
function acquires a mass term ${\rm e}^{-mr}$ with a 
non-vanishing mass in front of the scaling piece. 
The mass $m$ does not depend on $N$.
This means that the average distance (\ref{bpdH}) is of the order
$1/m$~:
\beq
\langle r \rangle \sim \frac{1}{m} \sim {\rm const} 
\eeq
and does not scale with $N$. This may be interpreted 
as an infinite Hausdorff dimension. This phase is called
a collapsed phase since it is dominated by short
branched polymers which do not grow, contrary 
to the generic situation dominated by the elongated 
polymers.

The collapse of the geometry is a result of the appearance of
singular vertices on branched polymers \cite{bb1,jk,bbj}. 
A singular vertex is a vertex with an order which grows extensively 
with $N$. The mechanism of the appearance
of the singular vertices can be described in terms of 
the balls-in-boxes model discussed in the next section. 
The results of the discussion are summarized in the table 
\ref{bptable}.
\begin{table}
\begin{center}
\begin{tabular}{| c | c | c |}
\hline
phase       & $\gamma$ & $d_H$ \\
\hline
generic     & $1/2$ & 2 \\
\hline
marginal($\beta\le3$) & $1/2$                 & $2$                   \\
($2<\beta<3$)  & $(\beta-2)/(\beta-1)$ & $(\beta-1)/(\beta-2)$ \\
\hline  
collapsed($\beta>2$)  & $2 - \beta$ & $\infty$ \\
\hline 
\end{tabular}
\end{center}
\caption{Critical exponents $\gamma$ and $d_H$ for
the three phases of the branched polymers model.}
\label{bptable} 
\end{table}
 
To end  this section let us briefly discuss 
the topological aspect of the model \cite{jk}. 
One may extend the
class of graphs to the ensemble of graphs with
loops. Such graphs can be generated by the perturbative
expansion of a zero dimensional field theory with 
a potential containing terms $\phi^q$. The coefficients
in front of the terms are related to the weights $p(q)$.
The tree diagrams discussed so far come from the 
leading term in the loop expansion corresponding to
the classical tree level. The number of diagrams with 
an arbitrary number of loops is not exponentially bounded
so that the entropy is not an extensive quantity and there 
is no thermodynamic limit. There are some ideas
how to cure the problem as discussed below \cite{bk1,ds,gm,jk}.

If one wants to sum 
over topologies in (\ref{bpz}) one has to consider 
topological terms in the action.
The number of loops $L$ is the simplest topological 
term for graphs. The partition function can be 
written as~: 
\beq
{\cal Z}(\hbar,\mu) = \sum_L \hbar^L Z_L(\mu)
\label{bpzz}
\eeq
where $Z_L$ is the partition for the sub-ensemble with
$L$ loops, and $\hbar$ is a coupling constant for
the topological term. The nonexistence of the 
exponential bound for the number of all diagrams 
implies that the series is not summable. It is not even
Borel summable. The radius of convergence is zero
and therefore there are functions which can be added
to ${\cal Z}$ without changing the coefficients $Z_L$ 
of the series. 
Such functions are called nonperturbative modes. 
In principle the number of nonperturbative modes 
is infinite. The idea is to reduce it as much as possible.
This is done by the loop expansion of the scalar field theory
generating the asymptotic series (\ref{bpzz}). This
series incorporates contributions from diagrams with 
any number of loops.  One can show that in the double scaling limit 
$\hbar \rightarrow 0 $ and 
$ t = \Delta \mu^{3/2}/\hbar=const$,
the susceptibility $\chi(t)$ is given by the Riccati 
equation in the scaling argument $t$. 
This equation uniquely determines all coefficients $Z_L$ of 
the series (\ref{bpzz}) and has only one non-perturbative 
parameter. Thus the number of perturbative modes gets reduced to 
only one and the goal of summing over topologies gets 
partially achieved. Unfortunately, so far no physical 
principle to fix this remaining free parameter is known.
The existence of the double scaling follows from the fact that the 
susceptibility exponent $\gamma$ grows linearly with the Euler number.
This was first discovered in two dimensional gravity \cite{bk1,ds,gm}. 
One also finds a linear dependence in the marginal and semi-generic
phases of branched polymers \cite{jk}.

\section{Balls-in-boxes model}
\setcounter{equation}{0}

In this section we discuss a model
of weighted integer partitions --  mean
field approximation for models of dynamical lattices.
The model undergoes a phase transition which has
many common features with the Bose-Einstein
condensation. The integer partitions of the model correspond 
to the partitions of vertex orders of dynamical lattices. 
The phase transition relies on a condensation which favours
partitions with one integer proportional
to the sum of all integers in the partition \cite{bbj}.
This integer corresponds to the singular vertex order 
on the random lattice \cite{hin,ckrt}.

The partition function of the balls-in-boxes 
model~:
\beq
Z(M,N)= \sum_{q_1,\ldots,q_M}p(q_1)\cdots
p(q_M)\delta(q_1+\cdots+q_M-N)
\label{bbzmn}
\eeq
describes weighted partitions of $N$ balls in $M$ boxes. 
The function $\delta()$ is the Kronecker delta.
The weight is a product of one-box weights $p(q_i)$ which
means that the numbers of balls $q$ in any two different 
boxes are independent of each other. 
The independence is weakly broken by the 
constraint on the total sum which prevents the factorization. 
This constraint makes the model nontrivial.
For the convenience we assume that each box contains at
least one ball $q\ge 1$. If we additionally choose $N=2M-1$
then the partition function (\ref{bbzmn}) is equal to the partition 
function (\ref{bpz}) of the branched polymer \cite{bbj,bb1}.
The numbers $q_i$ of (\ref{bbzmn}) correspond to 
the orders of vertices of the branched polymer. The Euler relation 
$\sum_i q_i = 2M-1$ introduces the constraint. 

In the large $M$ limit the partition function of the
model takes the form~:
\beq
Z(M,N)= e^{M f(\rho) + \dots}
\eeq
where $\rho = M/N$ is the average density of balls per
box, and $f(\rho)$ is the free energy density per box.
The model has a phase transition at a certain critical 
density $\rho_{cr}$ where the free energy is singular. 
The value of $\rho_{cr}$ depends on the choice of weights $p$'s.
In particular $\rho_{cr}$ may be moved away to infinity 
or to one. In either case the model would only have
one phase. Here we are interested in the situation 
where $\rho_{cr}$ is finite and the model has two phases
depending on whether $\rho$ is lager or smaller than 
$\rho_{cr}$. For example, for the weights~: 
\beq
p(q) = q^{-\beta}
\label{bbweights}
\eeq
the model undergoes the phase transition at 
\beq
\rho_{cr} = \frac{\zeta(\beta-1)}{\zeta(\beta)}
\label{bbcr1}
\eeq
where $\zeta$ is the Riemann Zeta function. To fix
attention and not to make formulas too abstract
we will keep in this section this particular form
(\ref{bbweights}) of weights, but the discussion
can be naturally generalized to other forms \cite{bbj,jk}. 
The critical density $\rho_{cr}$ is finite for $\beta$ 
larger than two. When $\beta$ goes to infinity the 
critical density approaches one.
When $\beta$ goes to two the critical density goes
to infinity and eventually the transition disappears 
when $\beta$ becomes equal or smaller than two.

An alternative way of triggering the transition is to
change $\beta$ for fixed $\rho$ as for instance for
branched polymers where density is fixed $\rho=2$.

The free energy density has the following singularity
at the phase transition~:
\beq
\partial_\rho f \sim \left\{ \begin{array}{lll}
\Delta \rho^{1/(\beta -2)} & \mbox{ for } & 2 < \beta < 3 \\
& & \\
\Delta \rho^{\beta - 2} & \mbox{ for } &   \beta > 3
\end{array} \right . \, .
\eeq
There are logarithmic corrections in $\Delta \rho$
for integer $\beta$. The free energy has the same 
type of singularity in the parameter $\Delta \beta$.

Analogously to branched polymers (see previous section),
an additional exponential factor in 
the weights (\ref{bbweights}) 
$p(q) = e^{\kappa q} q^{-\beta}$ does not affect the
phase structure \cite{bbj}.

It is convenient to consider the dressed one-box
probability $\pi(q)$ to see what happens in the system
at the transition. The dressed probability $\pi(q)$ is
a probability that a particular box has $q$ balls.
One obtains for large $M$~:
\beq
\pi(q) = \left\{ \begin{array}{lll}
\frac{q^{-\beta}e^{-\mu q}}{{\cal N}(\mu)} & \mbox{ for } & \rho < \rho_{cr} \\
&& \\
\frac{q^{-\beta}}{{\cal N}(0)} \, + \, \frac{1}{M} \delta(q - M(\rho-\rho_{cr}))
& \mbox{ for } & \rho \ge \rho_{cr}
\end{array} \right .
\label{bbpisp}
\eeq
where the normalization factor ${\cal N}(\mu)$ is~:
\beq
{\cal N}(\mu) = \sum_{q=1}^{\infty} q^{-\beta} e^{-\mu q} 
\eeq
and $\mu$ is a positive function of $\rho$
which vanishes at the transition. 
One can check that the average number of 
balls per box is indeed $\rho$~:
\beq
\sum_{q=1}^{\infty} q \pi(q) = \rho
\label{bbav}
\eeq
for the one--box probability $\pi(q)$ given by (\ref{bbpisp}). 
The interpretation of the 
result (\ref{bbpisp}) is following. In the low density phase 
($\rho<\rho_{cr}$),
the typical fluctuation of the box occupation 
number $q$ is of the order $1/\mu$. At the transition,
$\mu$ vanishes, so the fluctuations must be of
the order of the system size. Indeed, at the transition
one box captures a number of balls which grows with $M$ 
as shows the argument of the delta function in
the second term of the high density 
formula (\ref{bbpisp}), and the occupation of this singular box
gives rise to large fluctuations.
At the transition the probability $\pi$ has the critical form~:
\beq
\pi_{cr}(q) = \frac{q^{-\beta}}{{\cal N}(0)} \, .
\eeq
Above the transition this form is frozen but it is supplemented by
the anomalous term~:
\beq
\mbox{anomaly} = \frac{1}{M} \delta\Big(q - M(\rho-\rho_{cr})\Big) \, .
\eeq
This term is anomalous in the sense that it disappears from 
the probability distribution $\pi(q)$ if one takes 
the point-wise limit $M\rightarrow \infty$
for each fixed $q$. In this limit,
only $\pi_{cr}(q)$ survives. 
If one calculated the average (\ref{bbav}) for such a limiting probability 
distribution $\pi_{cr}(q)$, one would obtain the wrong value 
$\rho_{cr}$ instead 
of the correct one $\rho$. 
This means that the anomalous term can not be neglected in calculating 
the average (\ref{bbav}). The anomaly introduces an additional 
probability $1/M$ 
of picking one out of $M$ boxes with $M(\rho-\rho_{cr})$ balls. 
This is the singular box (singular vertex).
\begin{figure}[t]
\begin{center}
\epsfig{file=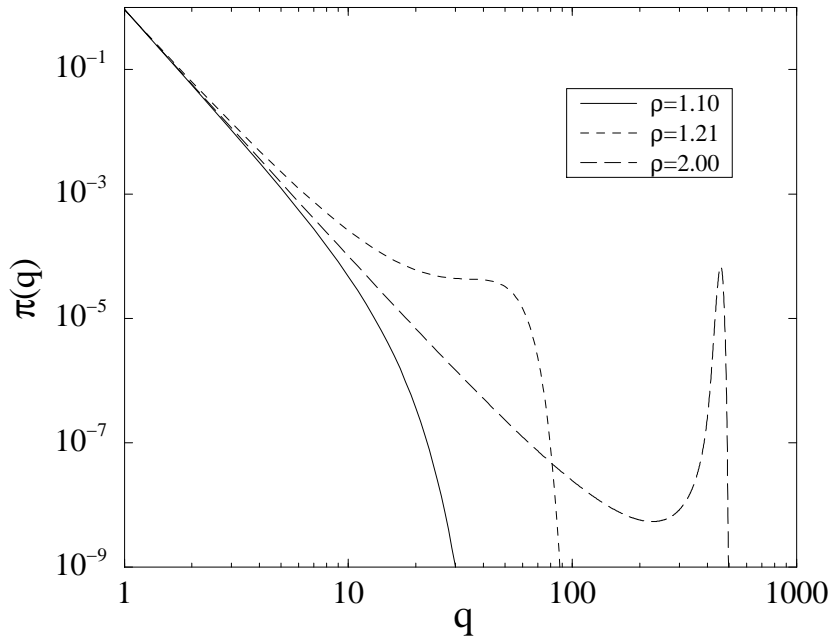,width=12cm,bbllx=12,bblly=40,bburx=576,bbury=460,clip=true}
\end{center}
\caption{\label{rbb1} Evolution of the shape of the
dressed one-box probability $\pi(q)$ with density $\rho$.
The three curves correspond to densities below, at  
and above the transition. (The model with the weights $q^{-\beta}$ 
for $\beta=4$) }
\end{figure}
As we see the anomaly corresponds to the condensation of balls in one
box which just takes over the surplus of balls holding the rest
of the system critical. The condensation is similar to the 
Bose-Einstein condensation. The difference between the two condensations
is that in the Bose--Einstein condensation, particles go to 
the lowest energy state, while here the box must be chosen 
by the symmetry breaking since the boxes are indistinguishable.

The transition to the condensed phase is visualized for the 
finite size system $M$ in the figure \ref{rbb1}. The finite 
size calculations have been done by an improved 
version of the recursive technique described in \cite{bbj}. 

\begin{figure}[t]
\begin{center}
\epsfig{file=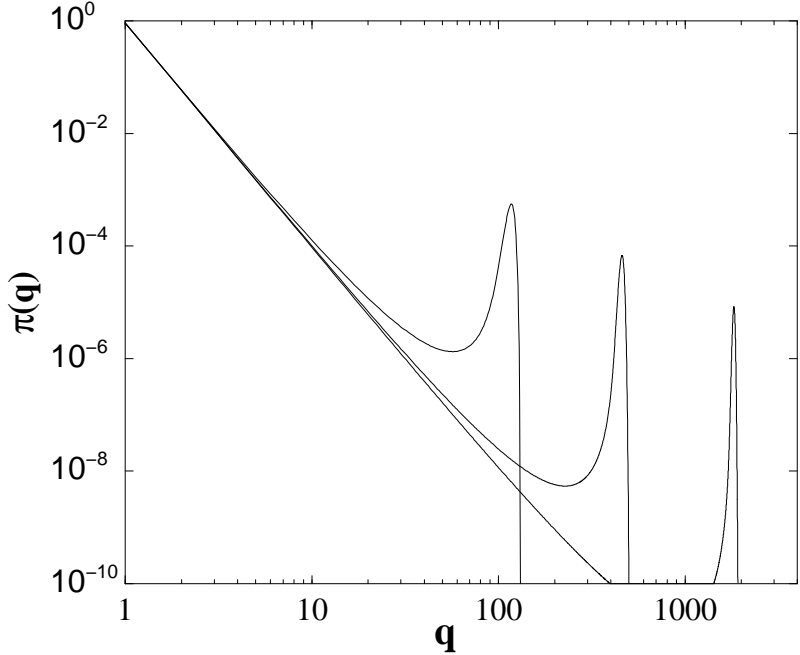,width=12cm,bbllx=12,bblly=40,bburx=576,bbury=460,clip=true}
\end{center}
\caption{\label{rbb2} The dressed one-box probability $\pi(q)$ in the
condensed phase for $\rho=2$, $\beta=4$, $N=128,512,2048$.}
\end{figure}

There are some secondary finite size effects to
the formula (\ref{bbpisp}) as for instance that
the peak at $M(\rho-\rho_{cr})$
is smeared for finite $M$ or that one has to 
go to sufficiently large $M$ to see the peak 
depart from the remaining part of the distribution. However,
the basic features of the solution (\ref{bbpisp})
that the position of the peak moves linearly with $M$ 
and that its height decreases as $1/M$ 
are already seen for moderate sizes $M$ (see figure \ref{rbb2}).

It is also interesting to consider ensembles with varying
density $\rho$ \cite{bb2,bbj2}. There are two natural candidates. The ensemble
with a variable number of balls with the partition function~:
\beq
Z(M,\mu) = \sum_{N} Z(M,N) e^{-\mu N}
\label{bbcan}
\eeq
or the ensemble with the variable number of boxes~:
\begin{equation}\label{zkappa}
Z(\kappa,N) = \sum_{M} Z(M,N) e^{\kappa M}
\label{bbccan}
\end{equation}
If the sum over $N$ in (\ref{bbcan}) extends to
infinity, the problem factorizes to $M$ copies 
of the urn-model \cite{rs,gs}.
It does not, however, if there is an upper limit 
$N_{max}$ for $N$. In this case the phase
structure is basically the same as in the $Z(M,N)$ 
ensemble. For large $\mu$ the system
is in the low density phase. 
At some critical value of $\mu$, the system enters the phase where one 
of the boxes captures the surplus of balls 
to maximize $N$. Analogously in 
the ensemble (\ref{bbccan}), for $\kappa$ above
a critical value, the system is in the high density 
phase realized by minimizing 
the number of boxes to the smallest available number $M_{min}$. 

\begin{figure}[t]
\begin{center}
\epsfig{file=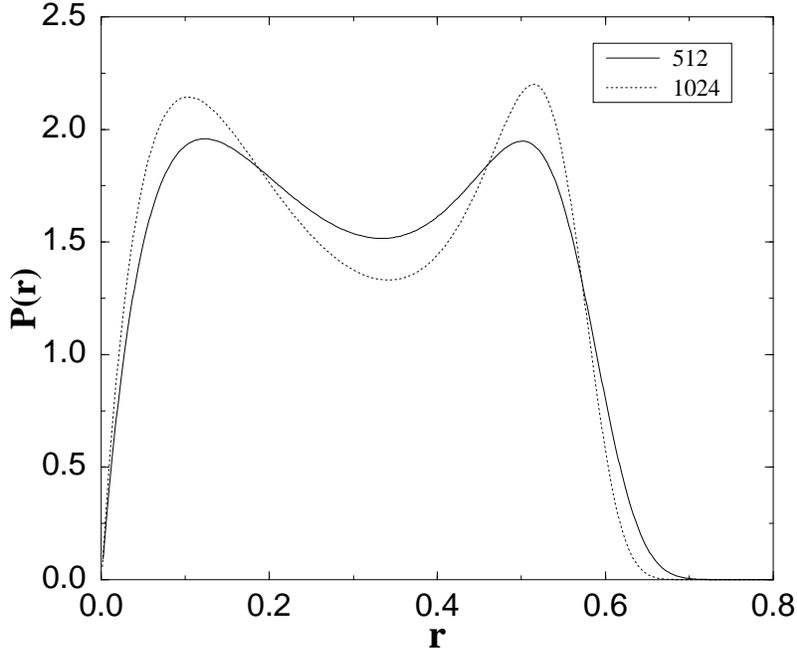,width=12cm,bbllx=12,bblly=40,bburx=576,bbury=460,clip=true}
\end{center}
\caption{\label{bb3} The distribution of $r=M/N$ in
the $(\kappa,N)$ ensemble in the pseudo-critical 
region for two different volumes~:
$(-0.32184,512)$, $(-0.31910,1024)$ for the model with
the weights $p(q)=q^{-\beta}$ for $\beta=2.5$.}
\end{figure}

If one considers the $(\kappa,N)$ ensemble,
it is more convenient to use the quantity~:
\beq
r = \frac{\langle M \rangle}{N} = 
\frac{1}{N} \frac{\partial Z(\kappa,N)}{\partial \kappa}
\label{ccan}
\eeq
instead of the balls density.
The values of $r$ are limited to the range~:
$M_{min}/N \le r \le M_{max}/N$. The lower
limit may be naturally chosen~: $M_{min}=1$,
to have at least one box, and the upper one~:
$M_{max}=N$, which corresponds to one ball per box. 
When $\kappa$ goes from large negative to large positive
values, $r$ goes from the lower to the upper limit.
In the large $N$ limit the function $r = r(\kappa)$ can
be found. For $\kappa<\kappa_{cr}$ the system stays at the lower
limit $r=0$. At the critical value it jumps to $r_0>0$
and then approaches continuously the upper limit when
$\kappa$ goes to infinity. The phase transition is discontinuous
and there is a latent heat related to the height $r_0$ of
the jump. 
For finite $N$ the transition between the phases is smoothed. 
There is a crossover interval of the size
$\delta \kappa \sim N^{-1}$ where the curve $r$
steeply goes between the two regimes. In this
crossover region system is effectively 
a mixture of two phases and hence the distribution
of $r$ has two peaks as shown in figure \ref{bb3}.
One peak is at the lower
kinematic limit while the other at $r_0$ in the large $r$ 
phase. When $\kappa$ moves in the crossover region,
the relative peak heights change.
One can define a pseudo critical value of $\kappa$ for a finite
system as the value at which the heights of both peaks become equal. 
We show in figure \ref{bb3} histograms obtained by finite 
size computations for two $N$'s.
The depth of the valley between the peaks increases
with the size since each of the peaks becomes
narrower. Eventually in the limit $N\rightarrow\infty$
the configurations from the valley are completely 
suppressed.

To summarize, the transition related to the appearance of
the surplus anomaly is continuous in the fixed density
ensemble and discontinuous in the $(M,\mu)$ or $(\kappa,N)$
ensembles with fluctuating density.

\section{Random surfaces}
\setcounter{equation}{0}
 
A theory of random surfaces has been an active research
field since Polyakov proposed the geometrical approach
to the quantization of strings by combining the Feynman 
quantization principle with the geometrical nature 
of the string action \cite{pb,p1}. There are some excellent 
reviews summarizing the models, the ideas and the methods 
\cite{d1,a1,gm2}. Here, for completeness, we briefly sketch 
the ideas and the main results which can be found there.
Then we discuss some issues which appeared 
later in the literature. 

Let us come to the origin. The partition 
function for the Polyakov theory reads~:
\begin{equation}
Z = \sum_{top} \int \frac{{\cal D} g_{ab}}{{\cal D} \mbox{{\small diff}}}
{\cal D} \varphi \exp\Big[ -S[g_{\mu\nu},\varphi]\Big] \,
\label{2dZ2}
\end{equation}
The action is a sum of the Einstein--Hilbert action (\ref{deeh}) 
and the action for the matter fields $\varphi$ coupled minimally to gravity.
The term~: $\int {\rm d}^2\xi \sqrt{g} R = 4\pi(1-h)$ 
of the Einstein--Hilbert action is a topological invariant 
and appears only if one sums over genera $h$.

The Nambu-Goto string embedded in $D$ dimensions 
rewritten in terms of the Polyakov formalism corresponds
to the action~:
\begin{equation}
S[g_{ab},\varphi] = \sum_{\mu=1}^{D} 
\int d^2 \xi \sqrt{g} g^{ab} \partial_a \varphi^\mu 
\partial_b \varphi^\mu + \mu \int d^2 \xi \,
\end{equation}
for $D$ scalar fields coupled minimally to gravity. 
The term proportional to the Euler characteristic is not
displayed.  One extends this action to non integer $D$.
In this case $D=c$ where $c$ is the central charge 
of the conformal matter coupled minimally to gravity.

As mentioned, the continuum Liouville field theory 
and the discretized  dynamical triangulation approach
used to calculate the partition function,
yield the same results for $c\le 1$. 
The approaches are independent. Even more, they are
independent to such a degree that it is very
difficult to find a direct correspondence between
them. It is much easier to compare results for the 
universal quantities than to compare 
the formalisms themselves. For example, a lot of
work has been spent to recover the complex structure 
or moduli spaces in the dynamical triangulation approach \cite{kty}.

The fundamental results of the theory are summarized
in the KPZ (Kni\-zh\-nik, Polyakov, Zamolodchikov)
formula \cite{kpz,d2,dk} for the scaling dimensions 
of operators dressed by gravity and the susceptibility
exponent $\gamma$ defined by the analogous formula 
as (\ref{bpzn}). Conformal weight $\Delta_0$ of an operator
acquires a new value $\Delta$ when the operator is coupled
to gravity.   The dressed value is given by the equation~:
\begin{equation}
\Delta - \Delta_0 = - \frac{\alpha^2}{2} \Delta (\Delta -1)
\label{kpz1}
\end{equation}
where 
\begin{equation}
\alpha = \frac{1}{2\sqrt{3}}(\sqrt{25-c} - \sqrt{1-c}) \, .
\end{equation}
The number of surfaces with area $A$ is given by
\begin{equation}
Z(A) \sim A^{\gamma - 3 } e^{\mu_0 A}
\end{equation}
where the entropy exponent $\gamma$ for spherical surfaces is~:
\begin{equation}
\gamma = \frac{1}{12}\Big(c-1 - \sqrt{(25-c)(1-c)}\Big) \, .
\label{kpz2}
\end{equation}
For other topologies the exponent $\gamma$ changes linearly 
with genus~: $\gamma_h = \gamma + h ( 2 - \gamma)$.
The Liouville theory breaks down 
at $c=1$. This is known as the $c=1$ barrier. 
This is related to the instability of the conformal
mode which drives the system to the branched polymer phase.
In the language of strings it corresponds to a tachyonic state
which destabilizes the stringy vacuum.

In the discretized approach, the functional
integral over surfaces is regularized by 
the sum over triangulations \cite{d3,kkm,adf}.
The discretized theory is given by the 
partition function~:
\beq
Z = \sum_{T} 
\prod_i q_i^{-\alpha} 
\frac{1}{C[T]} e^{-\mu A + \lambda h }
\exp \Big[ - \frac{1}{2} 
\sum_{\mu=1}^{D}\sum_{ij} (X^\mu_i - X^\mu_j)^2 \Big]
\label{aD}
\eeq
where the sum runs over triangulation, and $X^\mu$ are
$D$ scalar fields with the nearest neighbours interactions. 
For non integer $D$ one can either directly
weight triangulations by the power $-D/2$ of the determinant of the
Laplacian obtained by integration of the $X$ fields or
one can consider various statistical models corresponding to the
conformal matter field with $D=c$. The number of triangles 
on the lattice (area) is denoted by $A$, and the genus by $h$. 
The two terms $-\mu A + \lambda h$ correspond to the Einstein--Hilbert
action. 

The product of the powers of vertex orders
was originally introduced to investigate the stability 
of the discretized integration measure \cite{k1,bkkm,adf2}. 
The measure term corresponds to the higher derivative terms
and is irrelevant in the perturbative regime.  

\begin{figure}[t]
\begin{center}
\epsfig{file=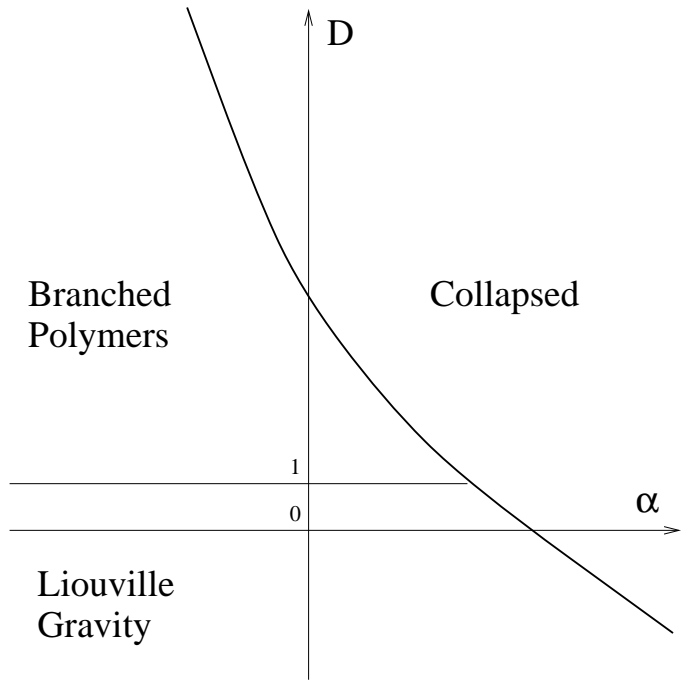,height=9cm}
\end{center}
\caption{\label{2d1} The phase structure of the model given
by the partition function (\ref{aD}) in the $(\alpha,D)$ plane.}
\end{figure}

The phase structure of the model in the $(\alpha, D=c)$ plane
has three phases~: 
the gravitational phase corresponding to the Liouville theory, 
the collapsed phase with singular geometries 
and the branched polymer phase \cite{bkkm,adf2,k2}. 
The phase structure is approximately sketched in figure \ref{2d1}.

For $c$ between zero and one, in the Liouville phase, 
there exists a discrete series of models 
of unitary conformal matter coupled to gravity 
with the conformal charge $c=1-6/m(m+1)$ 
which can be enumerated by an index $m=2,3,\dots$ \cite{d1,a1,gm2}. 
The susceptibility exponent for this series
is $\gamma=-1/m$ (\ref{kpz2}).
This series has a realization in terms of statistical
models on a dynamical triangulation. For example, the first member
of the series corresponds to the dynamical triangulation 
without dressing (pure gravity). The second to the critical
Ising spins on a dynamical triangulation, the third to the three-state 
Potts model. 
To associate with a statistical model a particular conformal 
field, one has to compare the operator contents. 
The conformal weights of the underlying conformal theory 
are related to critical indices of the corresponding 
statistical models. For example for the Ising model on random
lattice the values of the standard critical exponents 
\footnote{Compare with the Onsager exponents for the fixed lattice~:
$\alpha = 0$, $\beta = 1/8$, $\gamma = 7/4$,  $\delta = 15$,
$d \nu = 2$ } \cite{k2,bk4,bj1}~:
\beq
\alpha = -1 \, , \beta = 1/2 \, , \gamma = 2 \, ,  \delta = 5  
\, , d_H \nu = 3 \, .
\eeq
correspond to the exponents calculated from the 
conformal weights (\ref{kpz1}) of the $c=1/2$ conformal field 
dressed by the Liouville field.
The exponent $\nu$ appears in the combination
with the fractal dimension $d_H$,
which itself is a dynamical 
quantity. It will be discussed below.

Analytic calculations for the discretized random 
lattice are performed by the matrix model technique
(see reviews \cite{d1,a1,gm2}).
Let us briefly recall the idea. 
By the duality transformation one can rewrite 
the sum over triangulations as a sum over $\phi^3$ 
Feynman diagrams generated by 
the perturbation expansion of the $\phi^3$ matrix field theory 
in zero dimensions. Amongst diagrams 
of the $\phi^3$ theory
there are such which include tadpoles and self-energy 
sub-diagrams. These correspond to pathological 
triangulations containing for example triangles
whose two edges are glued together. 
One can remove tadpoles and self-energy sub-diagrams by using 
standard renormalization procedure for the perturbation theory. 
The renormalized theory has the same universal content encoded 
in the critical indices as the original theory.
In the Ising model, one can check this by 
direct calculations \cite{bj1}.
Universal properties do not change  when
instead of $\phi^3$ one considers the $\phi^4$ diagrams,
as expected on the grounds of the more general argument 
that the local properties of the lattice do not matter
in the limit when the lattice spacing goes to zero.
This intuitively means that in this limit one cannot 
distinguish whether the lattice is built from triangles, 
quadrangles or other polygons. 

The perturbation expansion of the matrix models generates
all terms in the partition function (\ref{aD}). 
The symmetry factor $1/C$ occurs automatically from 
the Wick theorem. The area term corresponds to the 
perturbation order of the diagram
which counts the number of vertices.
The topological term arises from the colour expansion
of the matrix field \cite{th}. 
The matter content of the theory is generated
by the multi matrix action with chain interactions \cite{k2,d4}.
In this way one can construct $c\le 1$ conformal matter field
from the unitary series $c=1-6/m(m+1)$ or some
non-unitary matter. Within the formalism one can calculate 
critical exponents, correlation functions, macroscopic 
loop amplitudes {\em etc}. The colour expansion of the matrix
model simultaneously incorporates contributions from 
all topologies. This leads to a partial solution 
of the problem of summing over genera by reducing the
number of nonperturbative modes similarly as discussed
in the section on branched polymers.
For $c=0$, for example, the sum over topologies is reduced
in the double scaling limit to solutions of 
the Painlev\'{e} II equation which has only two 
nonperturbative modes \cite{bk1,ds,gm}. 

Contrary to the Liouville phase, in the two other phases
the central charge does not determine the universality class 
of the model.
For example two different microscopic realizations 
of matter with the same large $c$~:
the multiple-spin model with $n=2c$ spin species 
and the Gaussian scalar model with $D=c$ fields,
have different susceptibility exponents $\gamma$ equal
$1/3$ and $1/2$ respectively. 
By using some general arguments one can show 
that there exists a possible series of models with 
positive values of $\gamma$ in the range $(0,1/2)$ \cite{d}.
We denote the value of the susceptibility exponent 
in this series by $\bar{\gamma}$. The models in 
the series are related to the unitary models 
$\gamma=-1/m$, $m=2,3\dots$ by~:
\beq
\bar{\gamma} = \frac{\gamma}{\gamma-1} \, .
\eeq
According to the picture advocated in \cite{d} such surfaces 
with $\bar{\gamma}$ look like trees of weakly touching bubbles
which themselves describe surfaces with $\gamma$.
The first model in this series $\bar{\gamma}=1/3$
corresponds to random surfaces consisting of bubbles
of pure gravity ($\gamma=-1/2$) weakly touching
each other. A microscopic realization of such a model
is the multi spin  model \cite{d}. In this model magnetized domains
cover the whole surface of the bubbles. Each bubble contains aligned
spins which decouple from geometry within the bubbles \cite{atw}.
Effectively each bubble behaves therefore as pure gravity. 
The domains of aligned spins try to minimize the mutual contact
border so  that neighbouring bubbles contact by a very narrow
neck. Such a surface indeed looks like a tree of bubbles.
\cite{w1,w2}.

A candidate for a model with $\bar{\gamma}=1/4$ ($m=3$) being
the next in the series should have bubbles with $\gamma=-1/3$
{\em ie} with the $c=1/2$ matter which is realized
by the Ising field or the Majorana fermions in continuum.
The exponent $\gamma$ consistent with $1/4$ was measured
at the phase transition of the model of random surfaces
with extrinsic curvature \cite{abjp}. Extrinsic curvature terms can be
obtained by integrating out fermions from the super-string 
theory \cite{w3} and therefore it is tempting to speculate 
that this model indeed inherits some fermionic properties 
leading to $\gamma=1/4$.

One can find a realization of the situation described above
in the matrix model by introducing a touching term  
\cite{k4}. Such a touching term allows
surfaces with a given $\gamma$ to touch each other.
It has a certain coupling constant, $x$, controlling the number 
of such touchings. The value of the exponent $\gamma$ 
as a function of the touching coupling $x$,
stays at the Liouville value $\gamma$ as long
as $x$ is smaller than a critical value $x_{cr}$. 
At the critical point $x_{cr}$ the value of $\gamma$ 
jumps to $\bar{\gamma}$, and then for $x>x_{cr}$,
to the branched polymer value $1/2$. The phase 
with $\bar{\gamma}$ is marginal in this model similarly
as the marginal phase in the model of branched polymers.  

The phase transition at $x_{cr}$ has been recently analyzed 
\cite{d6} in terms of the renormalization group flow in the 
parameter space $(x,\mu)$. There are two fixed points for $c<1$~:
one associated with the Liouville phase and the other
with the branched polymer phase 
and there is a multi-critical point related to them 
on the $x=0$ line. At $c=1$, the two fixed points merge 
at $x=0$, and for $c>1$ disappear from the real 
$(x,\mu)$ plane and become complex conjugate. 
When $c$ grows they slowly depart  
from the real plane. Renormalization group trajectories 
in the real plane that pass in the vicinity of 
the complex points look very much like 
in the limiting $c=1$ case, where $\gamma=0$. Only if one starts a
renormalization group trajectory at $\mu$ very 
close to the critical value $\mu_{cr}$ 
\begin{equation}
|\mu-\mu_{cr}| \sim \exp\Big[ - \frac{\rm{const}}{\sqrt{c-1}} \Big]
\end{equation}
one can avoid the critical slowing down from the trajectory passing
close to these complex points. One has to go to lattices
of the size $N \sim 1/|\mu-\mu_{cr}|$ to see the branched polymer 
value $1/2$. When $c$ is increased, the branched polymer regime comes
closer. This explains a long standing puzzle of 
values of $\gamma$ computed at finite volumes \cite{bkkm,djkp,adjt,kk}.
The values of $\gamma$, measured numerically, smoothly increase with 
$c$ and reach $1/2$ within the error bars only at $c$ around 5. 
The model with the touching term has been recently simulated numerically
\cite{pt} confirming the scenario of \cite{d6}.

This renormalization group picture furnishes completely our understanding
of the behaviour of the system at the border line between the 
Liouville phase and the branched polymer phase. Below we will discuss
the behaviour of the system at the border of collapsed phase. 
As we show, it 
can be understood in terms of the balls-in-boxes model.

With each triangulation one can associate 
a distribution of vertex orders $\{q_i\}$. 
The opposite statement is not true, since in general
it can be more than one triangulation associated
with a given distribution $\{ q_i\}$. Denote the number
of such triangulations by $N(\{q_i\})$.  The idea is \cite{bbpt} 
to substitute the sum over triangulations by a sum over 
the order distributions $\{ q_i \}$'s~:
\beq
\sum_T \frac{1}{C[T]} \dots \mapsto \sum_{\{ q_i \}} N(\{ q_i\}) \dots
\eeq
and to approximate $N$ by the mean--field formula~:
\beq
N(\{ q_i\}) \sim p(q_1)\dots p(q_N) \, ,
\eeq
where $p(q)$ are one--vertex terms.
The measure term $\prod q_i^{-\alpha}$ in the partition function (\ref{aD}) 
has the same factorized form in $q_i$'s, so it in a sense enhances  
the one--vertex terms in comparison with
multi--vertex terms neglected by the approximation. 
The matter fields also contribute to the one--vertex terms. 
In the large $D$ limit this contribution is approximately
$q_i^{-D/2}$ \cite{adfo}. 
For triangulations with $N$ vertices and  genus $h$
\beq
\sum_i q_i = 6N - 2 + 2h
\eeq
as results from the Euler relation. This constrains the
sum of the vertex orders. One obtains the following 
approximation of the partition function 
$(\ref{aD})$~:
\beq
Z \approx \sum_{\{q_i\}} P(q_1) \dots P(q_N) 
\delta(q_1 + \dots q_N - (6N-2+2h))
\eeq
where $P(q) = p(q) q^{-D/2 - \alpha}$.
One recognizes the balls--in--boxes model discussed 
in the previous section. Thus one expects that for large $D$ 
the transition to the collapsed phase, where singular vertices 
arise, occurs at the line $\alpha \sim -D/2$. Generally for 
any $D$, one may force the transition to the collapsed phase 
by taking $\alpha$ large enough.
Similarly to the collapsed phase of the branched polymer model, 
the number of triangles in the nearest neighbourhood of the singular 
vertex grows with total triangulation size. Following
this, the average distance between points does not 
grow with the triangulation size. Therefore
one can say that the Hausdorff dimension is equal infinity.

Let us discuss now the two other phases of the model.
A common feature of the Liouville gravity and the
branched polymers phase is that the typical surfaces in these
phases are highly branched. 
As a corollary of the KPZ relation the number of 
surfaces with the disc topology 
whose boundary is a minimal triangular loop is~:
\beq
N(A) \sim A^{\gamma-2} e^{\mu_{cr}A} \, ,
\label{opgf}
\eeq
where $\gamma$ is the susceptibility exponent for
spherical surfaces. A triangular loop on the spherical triangulation
with $A$ triangles divides the surface onto two discs with
the triangular loop boundary~: one with area $a$ and the other with
$A-a$.  The number of realizations of such a situation is \cite{jm}
\beq
M_A(a) \sim N(a) N(A-a) \, .
\label{MA}
\eeq
The smaller of the discs, looks as an outgrowth on the rest
of the triangulation. Such an outgrowth
is called a minbu~: {\bf mi}nimal {\bf n}eck 
{\bf b}aby {\bf u}niverse \cite{jm}. When one considers 
triangulations with fixed size $A$, one 
can skip terms independent of $a$ in the expression (\ref{MA}) 
since they give a normalization factor. What remains, is~:
\beq
M_A(a) \sim  \bigg(a ( A - a)\bigg)^{\gamma -2 } 
\sim \bigg(a ( 1 - a/A)\bigg)^{\gamma-2} \, .
\label{MA1}
\eeq
In the range $1 \ll a \ll A/2 $,  the distribution of minbu sizes scales 
as~: $M_A(a) \sim a^{\gamma -2}$. In this range the tree of minbus 
is self similar. For $\gamma>0$ the average minbu size~: $\int 
a M_A(a) \sim A^{\gamma}$ grows with the size of the triangulation.

There is a close relation between the branching structure of
the baby universes and the fractal structure of the random surface. 
The relation has been investigated 
by means of the real space renormalization group method \cite{jkk,bkk}.
The idea is to define an elementary blocking transformation
of the renormalization group using self similarity of the tree
of baby universes. This idea can be practically realized
by cutting off the last generation of minbus and calculating
a change of the scale associated with the rescaling of the
minbu tree. An interesting outcome of these studies is
that the change of the average distance between points 
$\langle r \rangle$ on the 
decimated triangulation is related to the change of scale~:
\beq
\frac{\langle r \rangle_2}{\langle r \rangle_1} \sim
\left(\frac{ \langle A \rangle_2}{A_1}\right)^{\nu}
\eeq
where the subscripts $1$ and $2$ refer to the ensembles
before and after the renormalization group transformation. 
The blocking is performed on the fixed area $A_1$ ensemble.
The exponent $\nu$ was found to approach
$\nu = 1/d_H=1/4$ for large lattices showing the
existence of a close relation between the fractal 
structure of random surfaces and the branching 
structure of baby universes.

One can determine the Hausdorff dimension directly 
from the scaling of the puncture-puncture correlation 
functions \cite{kkmw,aw}. The correlation function has been found 
analytically for pure gravity. The resulting form
shows indeed that it is a function 
of one universal parameter having the 
form $x=r/A^{1/d_H}$ where $d_H=4$. 

The fractal structure of random surfaces in the presence of 
matter is not yet fully understood \cite{aa1,aa2,b2}. 
There are different theoretical predictions for the value
of the Hausdorff dimension. The diffusion equation in the Liouville 
theory combined with the De-Witt short distance expansion 
of the heat kernel tells us \cite{k5,w4}
\beq
d_H = 2 \frac{\sqrt{25-c} + \sqrt{49-c}}{\sqrt{25-c} + \sqrt{1-c}} \, .
\label{dh}
\eeq
An alternative result obtained using the Hamiltonian
formalism where one identifies the geodesic distance with 
the proper time is \cite{ik}~:
\beq
d_H = \frac{24}{1-c + \sqrt{(25-c)(1-c)}}
\label{tm}
\eeq
Numerically the measurements of the matter fields with the central
charge $c$ in the range $0<c\le 1$ suggest that the Hausdorff
dimension $d_H$ is equal to four irrespectively of the matter dressing
\cite{aa2}. The are two ways of measuring the Hausdorff dimension 
numerically. Either one counts the number of lattice points $n(r)$ 
at a distance $r$ from a random point and averages it over points
and surfaces and then one fits the result to the formula
$n(r) \sim r^{d_H-1}$ \cite{kksw}. An alternative way 
is to measure the average distance $\langle r \rangle$ 
between all points on the lattice and then
determine $d_H$ from the scaling formula 
$\langle r \rangle \sim A^{1/d_H}$ (\ref{bpdH}) 
for large lattice sizes $A$ \cite{abbjp2}. 
For pure gravity $c=0$ and for $c=-2$, the two estimates give
the same values \cite{aaijkwy}. 
In general they need not be equal. The definition based 
on the average distance $\langle r \rangle$ is related 
to the universal scaling in $x=r/N^{1/d_H}$ and the 
mass exponent (\ref{mass}) and therefore 
it is closer in spirit to the continuum physics. 
For the numerical purposes the scaling argument 
is usually modified by a small finite shift 
$r\rightarrow r+\delta$~: $x = (r+a)/N^{1/d_H}$ which
can be neglected in the large $N$ limit \cite{aj2}.
Introducing the shift for finite $N$'s, improves
the fit quality in the whole range of $r$ \cite{aa1,aa2}. 

The results of very extensive simulations can be summarized
as follows.  The numerical measurements disagree with 
the transfer matrix prediction (\ref{tm}). 
For gravity in the presence of the matter fields
$0\le c\le 1$ the value of the Hausdorff dimension seems to
approximately equal four for all $c\le 1$. The values predicted
by (\ref{dh}) lie a bit outside the error bars of the measured
values, but contrary to this, for $c=-2$ one obtains a perfect 
numerical agreement $d_H = 3.574(8)$ with the diffusion formula 
(\ref{dh}) \cite{aaijkwy}. This shows that this part of the two 
dimensional theory is yet weakly understood. The fractal structure
is being currently intensively investigated.
 
Another quantity characterizing the fractal structure 
of random surfaces is the branching dimension $d_B$. 
It measures the scaling of the average number 
$n_0(r)$ of disconnected pieces of the ball's boundary 
with radius $r$~:
\beq
\langle n_0(r)\rangle \sim r^{d_B}
\eeq
The branching dimension was measured numerically for $c=1$ 
and was estimated to be larger that $2.5$ 
showing indeed a big rate of branching of the 
surface\footnote{The results of \cite{abbjp2} were obtained 
from fits without the shift $r \rightarrow r+\delta$.}  
\cite{abbjp2}.

The scaling dimensions $d_B$ and $d_H$ can be derived from the 
loop distribution function \cite{kkmw}. This distribution  
carries the most complete information about the fractal structure. 
The loop distribution function $\rho(r,L)$
is defined as the average number of loops of length $L$ 
on the boundary of ball with radius $r$. More precisely 
$\rho(r,L)dL$ is the average number of loops with lengths 
in the range $L$ to $L+dL$. 
The loop distribution was found analytically for pure 
gravity by the transfer matrix method \cite{kkmw}. 
For $A=\infty$ it reads~: $\rho(r,L) = 1/R^2 f(x)$, 
where $x=L/R^2$, and 
$f(x) = (x^{-5/2} + 1/2 \ x^{-1/2} + 14/3 \ x^{1/2})e^{-x}$. 
The moments of the distribution~:
\begin{equation}
\langle L^n \rangle = \int_{\epsilon}^\infty dL L^n \rho(r,L)
\end{equation}
are~:
\begin{equation}
\begin{array}{lll}
\langle L^0 \rangle & = c_0  r^3/\sqrt{\epsilon}^3 & \\   & & \\
\langle L^1 \rangle & = c_1  r^3/\sqrt{\epsilon} & \\  & & \\ 
\langle L^n \rangle & = c_n  r^{2n}  & \mbox{for  } n>1 
\end{array}
\label{ld}
\end{equation}
where $\epsilon$ is the short distance cut-off ({\em eg} lattice
spacing) and $c's$ are constants.
The zeroth moment corresponds to the number of loops of the ball's 
boundary and the first moment of the distribution corresponds to
the length of the boundary. The two scaling dimensions 
$d_H=4$, $d_B=3$ are related to the singular part of the loop 
distribution $\rho$ (\ref{ld}). 

To summarize this section. The two dimensional theory is in a
very good shape. The phase structure is determined. One understands
the behaviour of the system at the critical lines between phases. 
The Liouville phase, related to 2d quantum gravity, can be studied
by a variety of methods. For the time being, the only open question
is the fractal structure of surfaces in the Liouville phase.

\section{Monte Carlo simulations}
\setcounter{equation}{0}
 
A bonus from the lattice regularization is the
possibility to perform Monte  Carlo simulations.
Computer simulations provide a powerful experimental 
tool to investigate nonperturbatively
statistical systems. In many cases, where analytic 
techniques break down, computer simulations are 
the only method to study a model. This for instance is the
case for higher dimensional random 
geometries.

The basic idea behind the computer simulations
is to implement a Mar\-kov chain 
in the space of configurations with the
stationary distribution proportional to 
$e^{-S}$. The chain is determined
by the transition probability $p(1\rightarrow 2)$ 
between any two configurations $1$, $2$. In practice, one
proposes a simple scheme, called the elementary step of the algorithm,
of modifying a current configuration to obtain its successor 
in the chain. One can show that the detailed-balance
condition imposed on the transition probabilities~:
\beq
e^{-S(1)} p(1\rightarrow2) = e^{-S(2)} p(2\rightarrow 1)
\eeq
and the ergodicity of the elementary steps suffice
for the algorithm to generate configurations from
the stationary distribution $e^{-S}$.

\begin{figure}[t]
\begin{center}
\epsfig{file=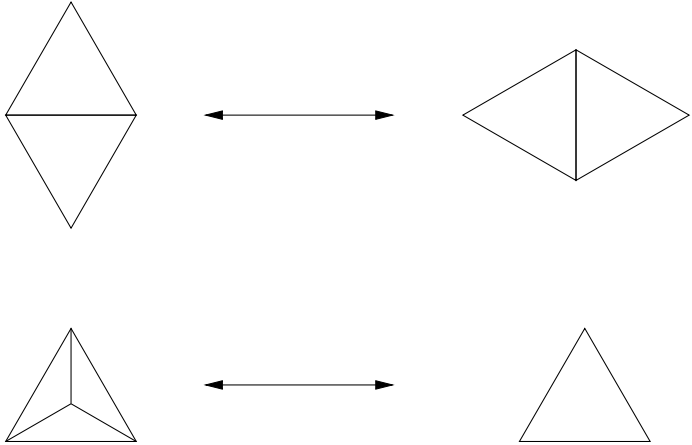,height=7cm}
\end{center}
\caption{\label{mc1} Ergodic set of local transformations 
of dynamical triangulations}
\end{figure}

An ergodic set of local operations to simulate two
dimensional dynamical triangulations is shown in figure \ref{mc1}.
The flip operation preserves the area. It is ergodic 
in the ensemble of triangulations with a fixed area \cite{bkkm,a}. 
The other two operations in the figure
which remove or add a point 
of order three, change the area and allow for 
extending simulations to the grand-canonical ensemble. 
The three moves form a set of moves called in general
the $(p,q)$ moves \cite{gv}. The first argument $p$ corresponds to the 
number of triangles before the transformation, and $q$ after it.
Note, that the four triangles, $p+q=4$, form a tetrahedron when 
glued together \cite{bk2}. This observation was used 
to generalize these transformations to higher dimensions, 
as we shall see later.
There is also another ergodic set 
of transformations, called the split and joint operations, also
used in update schemes \cite{jkp}. The two sets are equivalent.

The standard algorithm was extensively tested 
in two dimensions. The distribution of triangulations 
generated by the algorithm is in 
perfect agreement with the analytic formula for
the diagram enumeration.  Numerical measurements of the critical 
exponents of the statistical models $0< c \le 1$ are 
in excellent agreement with the KPZ results \cite{bj2}. 
The agreement extends beyond the critical 
region as is shown by the comparison of the numerical results for
the Ising model on the dynamical triangulations and the 
analytic results of the two matrix model \cite{bj1,bjk}. 
This can be treated as a proof for the practical 
ergodicity of the Monte Carlo algorithms. This practical proof is in a sense 
stronger than the mathematical proof which only states
the existence of a Markov chain between any two configurations which
may of course not suffice for practical purposes.

The $(p,q)$ moves have a natural generalization to higher
dimensional simplicial manifolds \cite{gv}. In four dimensional 
case there are five moves $p+q=6$, $p=1,\dots,5$. Geometrically
they can be viewed as a substitution of $p$ 4-simplices 
from the triangulation by  $q$ new simplices being 
a complementary part of the boundary of a 5-simplex.
One can show that the $(p,q)$ moves are equivalent
to the Alexander transformations \cite{a} known to be ergodic in the set of
combinatorially equivalent simplicial manifolds with a fixed
topology. As in two dimensions, one can show 
the practical ergodicity of the algorithm \cite{bbkp2}. 
One has to keep in mind, however, as prompted in \cite{bn}, that
the recognizability conjecture states that for some topologies 
a fraction of manifolds accessible by a Markov chain may scale 
to a number less than one in the large volume limit.
This may systematically bias the analysis of the finite size scaling.

The ergodicity and the detailed balance condition 
leave a large freedom for the invention of optimal 
algorithms. The local update schemes are known in general
to suffer from the slowing down, decreasing the 
algorithmic efficiency. The reason lies in the
random--walk nature of local changes -- namely,
many changes are undone by successive steps of the 
algorithm. A general strategy to cure the problem is to  
implement algorithms focusing directly on physically important 
modes. This strategy has been frequently used in the standard
field theoretical Monte Carlo simulations as for instance 
in the multi scale or cluster algorithms \cite{sw,w5}. 

\begin{figure}[t]
\begin{center}
\epsfig{file=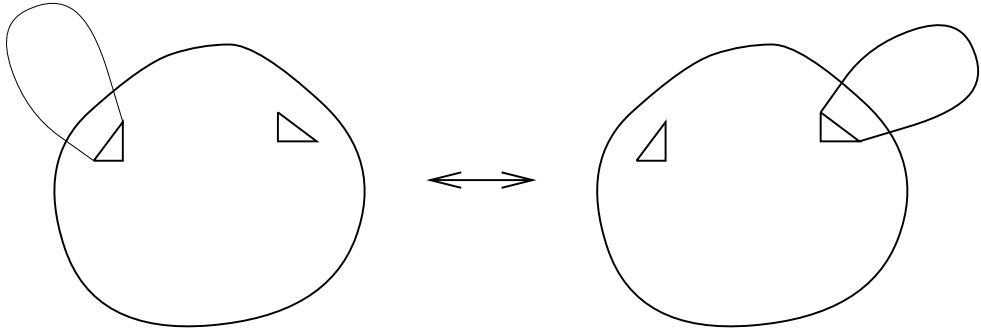,width=13cm}
\end{center}
\caption{\label{mc2} Elementary step of the baby universe surgery on
dynamical triangulations.}
\end{figure}
As discussed, typical random surfaces from the branched polymer phase 
or from the Liouville phase are populated with baby universes 
forming self similar trees \cite{jm}. The idea to use the 
tree structure in the update scheme leads to 
the baby universe surgery algorithm \cite{abbjp3} .
The algorithm is in fact very simple as depicted
in figure \ref{mc2}. One finds a minimal neck on the surface, 
cuts the surface along the neck, removes the corresponding 
minbu and pastes it into a randomly 
chosen place on the rest of the surface. Reshuffling minbus speeds up the 
algorithm dynamics by decorrelating the tree branches. 

A typical quantity characterizing the efficiency is
the autocorrelation time. It tells us, roughly speaking,  
how many sweeps of the algorithm are needed to decorrelate
the measurements of a given observable. The rise of the 
integrated autocorrelation time $\tau$ for large system 
sizes $A$ is controlled by the dynamical critical exponent $z$~: 
\beq
\tau \sim A^z
\eeq
In the table \ref{mctable} we present the comparison of values of the
dynamical exponent $z$ for the standard algorithm and a hybrid
of the standard algorithm with the minbu surgery for various
observables measured in the model with one scalar field. 
The values of the exponent $z$ get generally reduced 
if one supports the algorithm by the minbu surgery \cite{abbjp3}.
This improves significantly the algorithm dynamics.
The baby universe surgery is also applied to simulate 
higher dimensional simplicial manifolds in the elongated phase 
\cite{aj2,bkk}.
\begin{table}
\begin{center}
\begin{tabular}{| c | c | c |}
\hline
quantity & $z_{local}$ & $z_{hybrid}$ \\
\hline
$d$      &  1.06(3) & 0.76(3) \\
$d_{xy}$ &  0.81(6) & 0.14(2) \\
$r$      &  1.4(1)  & 0.50(3) \\
\hline
\end{tabular}
\end{center}
\caption{The dynamical exponent for the standard local
algorithm and the hybrid with the baby universe surgery,
for the simulations of surfaces with $c=1$, for 
the averaged internal distance between points~: $d$, 
the distance between two given points~: $d_{xy}$
and the gyration radius~: $r$. One sees a large 
reduction of the exponent $z$ when one adds the surgery to the update
scheme.}
\label{mctable}
\end{table}

Apart from the dynamical Monte Carlo techniques described above,
in some particular cases there are the so-called static algorithms.
They sample directly the static distribution $e^{-S}$ without 
using an auxiliary Markov chain. 
Thus, they are free of the dynamical slowing down.
Example of such an algorithm is a recursive sampling technique,
applicable for $c=0,-2$. In these cases there are known analytic 
formulas for the diagram enumeration which allow one to 
directly construct the diagrams weighted by the static 
distribution $e^{-S}$ \cite{kksw,am4}. 

The numerical algorithms for sampling of random geometries have
become a well established method allowing for studying 
triangulations with sizes ranging up to million triangles 
or hundreds of thousands of 4-simplices in the four dimensional case.

\section{Simplicial gravity}
\setcounter{equation}{0}

Simplicial  gravity is an attempt to formulate
the quantum theory of gravity. It is a part of 
a larger programme based on the assumption 
that one can apply the same set of 
fundamental principles as in field theory, 
to quantize gravity. 
Following these lines one  
tries to apply the Euclidean version of the
Feynman formalism as already described for
the two dimensional gravity (\ref{2dZ2}). 
There are many  conceptual problems related to 
the Euclidean formulation like for example
the lack of formal conditions 
which would ensure that 
we can reconstruct the Minkowskian quantum gravity.
This is, however, a part of a more general
difficulty, namely that we do not know how to formulate the
Minkowskian quantum gravity, so we do not know what 
exactly should be reconstructed. The role of the topology is also
not clear. And again, the problem is more general 
since we can not classify four dimensional topologies. 
Finally, there is a problem with the 
unboundedness of the action coming from the conformal mode. 
This problem, fortunately, is automatically cured 
by the discretization scheme. Having all these difficulties
in mind let us formulate in the beginning a more modest
aim, to define consistently the Feynman integral 
over the Riemannian structures with a fixed topology.
Now the measure problem arises. It is much 
more pronounced in four dimensions than it is in two.
One way to bypass it  is to formulate the 
theory perturbatively in terms of the Gaussian 
measure which is well defined. The 
perturbation theory obtained in this way is however
not renormalizable and the treatment fails.
One can improve the convergence of the perturbative diagrams
at large momenta by resummation techniques but then
one encounters problems with unitarity \cite{s}. 

One attempt of the nonperturbative formulation of quantum gravity is
based on the $2 + \epsilon$ expansion \cite{w6,kn,ak}. 
The situation is somewhat similar to 
the nonlinear sigma model \cite{bzg} where the perturbative 
treatment does not work above two dimensions but one can find 
a well defined nonperturbative fixed point in terms of the 
$2+\epsilon$ expansion. The $2 + \epsilon$ expansion 
may in principle be applied to gravity but in this case one
has two conceptual difficulties. First of all, $\epsilon$ 
equal two is not a small parameter, and second of all,
two dimensional integrals appearing as coefficients in the 
$\epsilon$-expansion are not able to reproduce the whole
content of higher dimensional gravity.

Another way to define the Feynman integrals
nonperturbatively is provided by
the lattice regularization.
In this context it was proposed in \cite{am1,aj1}.
This is a generalization of the dynamical triangulation approach 
from two \cite{d3,kkm,adf} and three dimensions \cite{am2,bk2,av}.

A lattice formulation allows the use of the
statistical ideas and techniques to calculate
quantum amplitudes which in the statistical
language correspond to some averages over the
statistical ensembles. The standard procedure
of the the statistical approach is well established.
Namely, given a model one addresses the following 
questions~:
\begin{enumerate}
\item does the model posses 
a well defined thermodynamic limit,
\item what is the phase structure of the model,
\item can one define a lattice independent 
continuum limit. 
\end{enumerate}
The last issue is related to the existence
of a continuous phase transition and the infinite
correlation length of some physical excitations
which would be independent of the short-range details 
of the lattice.

Let us follow these lines in the presentation.
We consider an ensemble of simplicial manifolds 
with a fixed topology. Four dimensional simplicial 
manifold consists of equilateral 4-simplices glued 
together in such a way that each two neighbouring 
simplices share a three dimensional 3-simplex.
The neighbourhood of each vertex is homomorphic 
to the $4$-ball. 

As discussed in the previous sections,
the sum over one dimensional graphs could 
be generated by the perturbative expansion 
of the scalar field theory while the sum over
two dimensional graphs by the matrix field theory. 
The extension of this idea to the tensor model to 
generate four dimensional simplicial manifolds does not work,
though it might seem to be an evident generalization
at a first glance.  The reason is that the perturbation expansion of  
tensor models generates graphs with fluctuating topology.
Even worse, the topology of diagrams fluctuates locally 
so that such diagrams do not correspond to manifolds \cite{gm2}. 
For the time being the only method to investigate
the sum over the ensemble of four dimensional simplicial manifolds 
are the  numerical simulations combined with the standard
statistical data analysis.

One considers the grand-canonical ensemble of 
simplicial manifolds with a given topology.
The partition function reads~:
\beq
{\cal Z}(\kappa_4,\kappa_0) = \sum_{T} \frac{1}{C[T]} 
e^{\kappa_0 N_0[T] - \kappa_4 N_4[T]} 
\eeq
The discretized Einstein-Hilbert action (\ref{deda}) 
reproduces in the naive continuum limit
the continuum counterpart (\ref{deeh}).
For convenience we substituted the number of
triangles $N_2$ from the formula (\ref{deda})
by the number of vertices $N_0$. This can be done
since the numbers $N_i$ of $i$-simplices on the
manifold are linearly related by the 
Euler and Dehn-Somerville
relations, which leave only two independent $N_i$'s. 
In particular, for the four dimensional sphere 
$N_0$ and $N_2$ are related by
$N_2 = 2(N_0 + N_4 -2)$.

One can rewrite the grand-canonical 
partition function as a sum~: 
\beq
{\cal Z}(\kappa_4,\kappa_0) = \sum_{N_4} Z(N_4,\kappa_0) e^{-\kappa_4 N_4}
\eeq
where $Z(N_4,\kappa_0)$ are the partition functions
for the canonical ensembles which have a fixed
volume $N_4$. One can go one step further
and express $Z(N_4,\kappa_0)$ in terms of
the state density $z(N_4,N_0)$~:
\beq
Z(N_4,\kappa_0) = \sum_{N_4} e^{\kappa_0 N_0} z(N_4,N_0) 
\label{nk}
\eeq
The thermodynamic limit exists if the
free energy density has a well defined 
large $N_4$ limit. This means that 
the function $\Delta$ in the formula
\beq
\log Z(N_4,\kappa_0) = N_4 \{ f(\kappa_0) + \Delta(N_4,\kappa_0) \}
\eeq
must be a finite size correction $\Delta$ 
which vanishes for large volumes 
$N_4\rightarrow \infty$~: 
\beq
\Delta(N_4,\kappa_0) \rightarrow 0.
\label{delta}
\eeq
so that in the limit $N_4 \rightarrow \infty$, the free
energy density $\log Z(N_4,\kappa_0) /N_4$ is a function
of $\kappa_0$ only.

One can show that if the function $\Delta$ vanishes for one 
finite value of $\kappa_0$ it does so for all other. 
The function 
$\Delta$ was studied for different values 
of the coupling $\kappa_0$ \cite{db2,aj3,bm2,bbp}. In particular 
it was found that for $\kappa_0=0$ the function delta scales as 
$\Delta \sim N_4^x$ where $x = 0.5(2)$ 
for manifolds with spherical and toroidal topology \cite{bbp}.
This is a numerical proof for the existence of
the thermodynamic limit. For $\kappa_0=0$
the partition function $Z(N_4,\kappa_0=0)$
is a sum of all manifolds without an extra weight.
This sum is an object of intensive mathematical studies 
\cite{dj,cm}. The goal of these studies is to prove
the existence of the exponential bound $Z(N_4,\kappa_0=0)\le  e^{c N_4}$.
Of course, this is equivalent to the existence of the thermodynamic 
limit of the model. So far there is no such  proof 
and one has to rely on the numerical results. 

One can show, as we shall see later,
that the pseudo-critical value $\kappa_{0,cr}$, at which 
the system enters the generic branched polymer phase 
is finite, or more precisely, it is bounded from
above by a finite value independent of $N_4$. On the 
other hand, it is known 
that the branched polymer phase has a well defined 
thermodynamic limit.  Combining these two facts, one
concludes that there exists a finite value of $\kappa_0$, 
for which the equation (\ref{delta}) is fulfilled. This 
suffices to end the proof of existence of the thermodynamic
limit. In our opinion this is the strongest numerical evidence 
for the existence of the thermodynamic limit.
What is nontrivial here is that
the pseudo-critical value $\kappa_{0,cr}$ does not move to
infinity when $N_4$ is increased as would happen if 
there was no thermodynamic limit.
 
Numerical simulations were performed for three 
topologies~: the sphere $S^4$ and tori 
$S^1\times S^1\times S^1\times S^1$ , $S^3\times S^1$.
In all these cases the free energy density $f(\kappa_0)$
was found numerically to have the same 
thermodynamic limit \cite{bbp}. 

Let us now outline the basic facts gathered 
using the computer simulations about
the phase structure of the model.

\begin{figure}[t]
\begin{center}
\epsfig{file=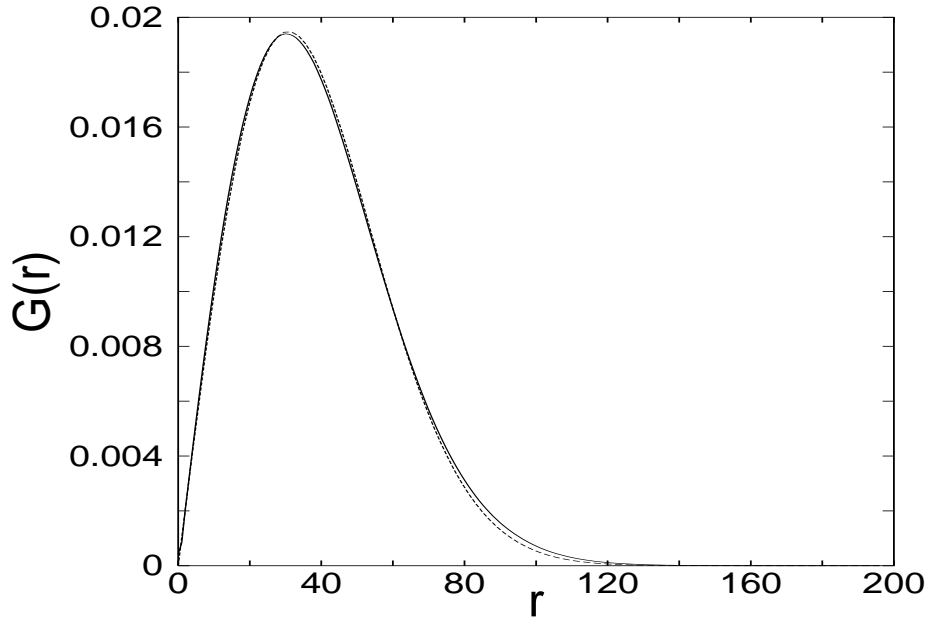,width=12cm,height=8cm}
\end{center}
\caption{\label{4d2} The minbu-minbu correlation function measured
on minbu trees in the elongated phase of simplicial
gravity (dashed line), compared with 
the universal formula $G(r) \sim 2ar e^{-ar^2/N}$ for the branched polymers 
(solid line) (\ref{bpg1}).}
\end{figure}

Simplicial gravity has two phases,
crumpled and elongated,  named to
reflect their geometrical properties.
The elongated phase corresponds essentially
to the branched polymer phase \cite{aj2}. In this phase
typical simplicial manifolds are 
populated with
baby universes which form the generation trees. 
The susceptibility exponent is $\gamma=1/2$.
The puncture-puncture correlation function is 
in this phase given by the universal formula 
(\ref{bpg1}), (\ref{bpg2}) for branched polymers.
\begin{figure}
\begin{center}
\hbox{ \epsfig{file=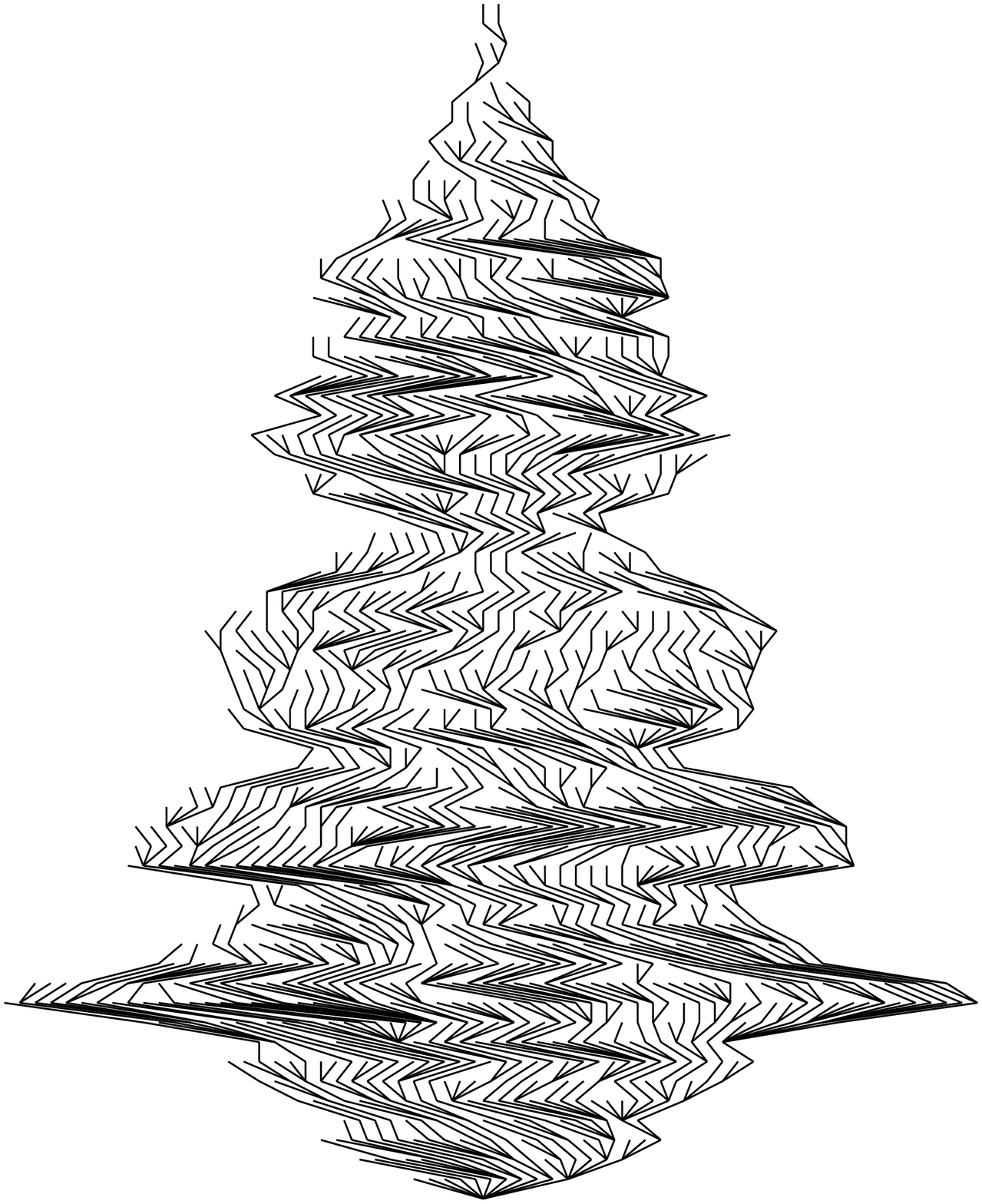,width=8cm}
\epsfig{file=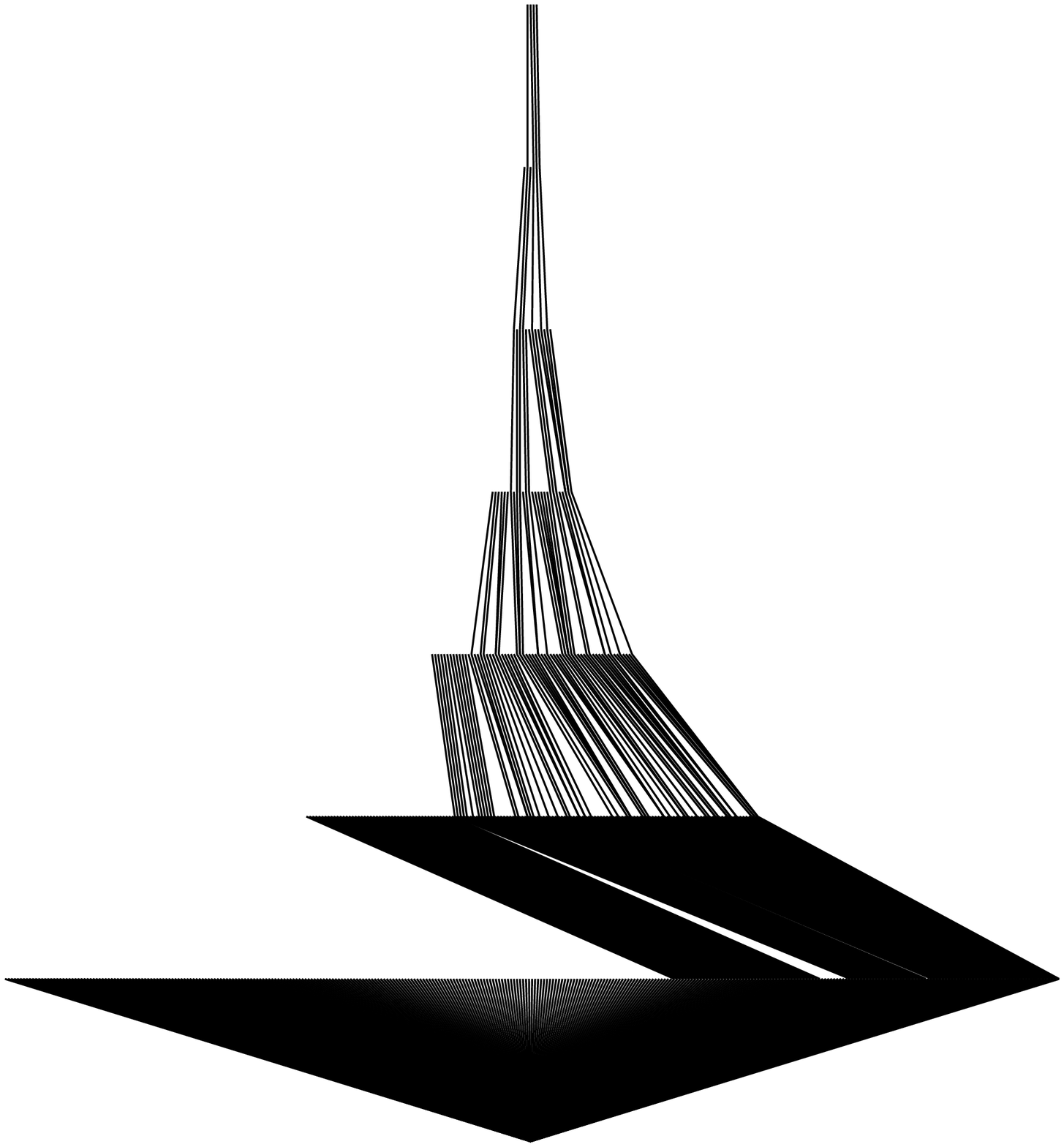,width=4cm} \hspace{2.0cm}  }
\end{center}
\caption{\label{4d1} Minbu trees for configurations
taken at random from the elongated phase (left)
and the crumpled phase (right). Each link on the tree
corresponds to a minimal neck on the simplicial manifold, and 
the end points of the link correspond to two parts of the simplicial
manifold on both sides of the minimal neck. The number of links
emerging from a vertex of the tree corresponds to the number
of minimal necks found directly on the part of simplicial
manifold associated with the vertex. The trees carry only
topological information about the connectivity of minbus.
The vertical ordering seen on the picture results from
visualization procedure and has no relevance 
for the tree structure. The tree on the left hand side 
has many generations and no distinguishable points.
The tree on the right hand size is short and has one 
singular vertex drawn as a root of the tree. There are
so many links emerging directly from the singular vertex
that the visualization procedure failed to draw
them as separate lines and instead drew a densely 
covered triangle.}
\end{figure}
The only dependence on $\kappa_0$ enters the formula through
one universal parameter $a$ as in the formula (\ref{bpg1}). 
In the figure \ref{4d1} we show the minbu trees constructed on simplicial
manifolds picked up randomly from 
two different phases of simplicial gravity.
The tree on the left hand side comes from the branched polymer
phase.  A vertex on the tree corresponds to a minbu
and a link to the minimal neck between neighbouring 
minbus. One can define  a distance between vertices on such 
a tree as a number of links between them. Then 
one can measure a minbu-minbu correlation function 
analogous to the puncture-puncture correlation 
function \cite{bbpt}. The results fit perfectly  the branched 
polymer correlation function (\ref{bpg1}) as shown 
in figure \ref{4d2}. This means that the minbu trees are
indeed branched polymers. 

\begin{figure}[t]
\begin{center}
\epsfig{file=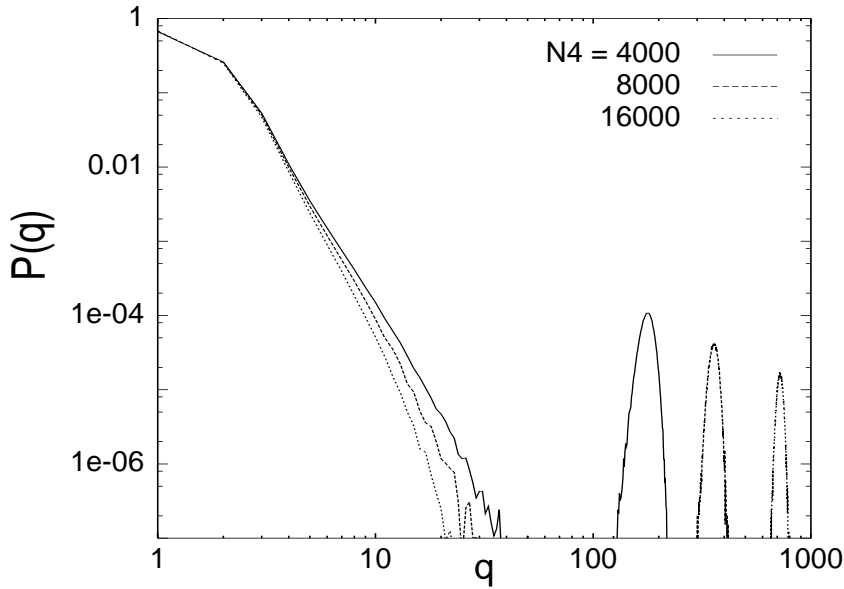,width=12cm,height=8cm,bbllx=40,bblly=20,bburx=600,
bbury=780}
\end{center}
\caption{\label{4d3} The distribution of the minbu orders in the crumpled 
phase of simplicial gravity at $k0=1.0$ 
for three different $N_4=4000,8000,16000$. The position of the 
peak corresponding to the singular minbu vertex shifting
proportionally to $N_4$.}
\end{figure}

The branching structure of minbus collapses when  the system enters
the crumpled phase. The collapse is related to
the appearance of a singular vertex on the minbu tree \cite{hin,ckrt}. 
On the tree on the right hand side
in figure \ref{4d1},
the tree has only few generations and 
one vertex has many branches. This is 
the singular vertex. One can check
that the number of branches emerging from
 the singular vertex
grows with the total number of minbus. This is
shown in figure \ref{4d3}.
To distinguish this minbu from the rest, one calls
it the mother universe. The volume of the mother 
universe grows extensively with the total 
volume of the simplicial manifold \cite{bbpt}. 
The reason for this rapid growth is related to the singular
vertices residing on the mother universe. 
There are two singular vertices at a 
distance one. They form a singular link \cite{ckrt}. 
The volume of the neighbourhood of the singular
vertices grows proportionally to the total volume. 
The situation is analogous to collapsed
branched polymers. Simplicial manifolds have 
infinite Hausdorff dimension in this phase.

Let us finish the survey of properties
of the model by describing what happens
at the phase transition. 
The standard way of investigating the behaviour
of a model at a phase transition 
is to perform the finite size analysis 
of quantities related to derivatives of the free energy. 
For the transition driven by $\kappa_0$ 
one investigates the second cumulant~:
\beq
c_2(\kappa_0,N_4) = 
\frac{1}{N_4}\frac{\partial^2 F(N_4,\kappa_0)}{\partial \kappa_0^2} = 
\frac{\langle N_0^2 \rangle - \langle N_0 \rangle^2}{N_4}
\label{c2} 
\eeq
which has an interpretation of the heat capacity and is
a measure of thermodynamic fluctuations. The cumulant
is related to the integrated correlation function 
of the curvature so its behaviour indirectly measures 
the signal from the two point function \cite{bbkp,dbs2}. In particular,
if the transition is second order, this signal may
be related to the occurrence of long range correlations
in the system.

In the thermodynamic limit, $N_4 \rightarrow \infty$,
the fluctuations are expected to approach a 
$N_4$--independent value $c_2(\kappa_0)$ 
except at the transition point where the large
$N_4$ behaviour is approximately  given by 
the finite size scaling formula~:
\beq
c_2(\kappa_0,N_4) \sim N_4^{\alpha/{d_H \nu}} 
f\big( (\kappa_0 - \kappa_{0,cr}) N_4^{1/d_H\nu}\big)
\label{scaling}
\eeq
where $\alpha$ and $\nu$ are the standard critical 
indices. They are related by the Fisher scaling relation
$\alpha = 2 - d_H\nu$. 
The value of the product $d_H\nu$ is bounded
by the thermodynamic inequality~: $1 \le d_H\nu$. 
In the limiting case $d_H \nu =1$, the transition is of first order.
In this case the exponent $\alpha=1$. The prefactor in the scaling 
formula (\ref{scaling}) grows linearly with $N_4$.
When $1<d_H\nu < 2$ and $0 < \alpha < 1$,
the transition is of second order. The prefactor in (\ref{scaling}) 
grows as a power of $N_4$, between zero and one. Finally, 
when  $d_H\nu > 2$ and $\alpha<0$, the prefactor 
in (\ref{scaling}) does not grow with $N_4$ and then 
$c_2$ approaches a finite constant at the transition. 

\begin{figure}[t]
\begin{center}
\epsfig{file=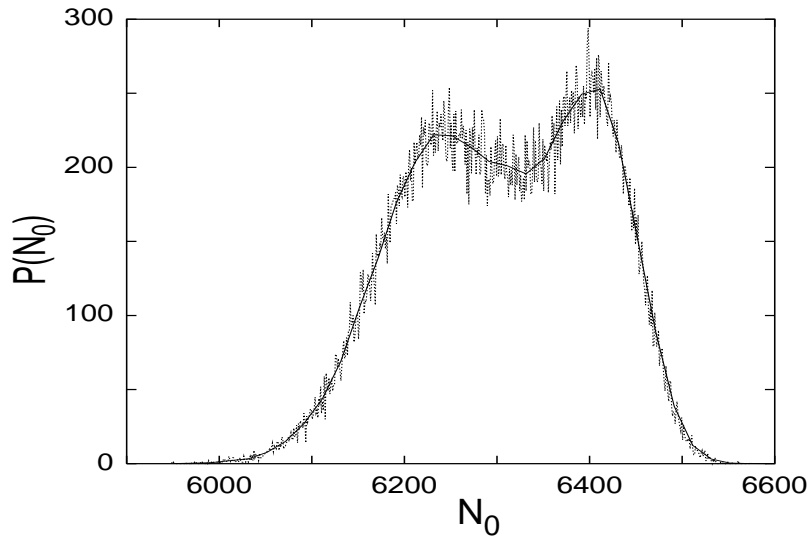,width=12cm,height=8cm,
bbllx=20,bblly=120,bburx=600,bbury=600,clip=true}
\end{center}
\caption{\label{4d5} The distribution of $N_0$ at $\kappa_0=1.258$ for 
$N_4 = 32000$ measured in the computer simulations. The vertical
axis corresponds to the number of entries for a given $N_0$.}
\end{figure}

Finite size analysis of the second cumulant shows 
that the large $N_4$ behaviour is in agreement
with the first order scaling \cite{bbkp}.
The observed linear rise with the volume
of the maximum of the heat capacity is related to 
the latent heat. Hence one expects a double peak
structure in the $N_0/N_4$ distribution. Indeed, 
such a structure has been found (see figure \ref{4d5})
at some pseudo-critical value of $\kappa_0$ \cite{bbkp,db1}.

To summarize, there are two phases separated by
a first order phase transition. This means that
there is no continuum theory associated with 
this critical point. We shall discuss the issue of the
continuum theory at the end of the section. 

Before doing this let us come back to the 
thermodynamics. The scaling properties of the 
elongated phase are exactly the same as those 
of the generic branched polymers as shows the
analysis of the minbu-minbu
correlations. The analysis
of the minbu trees suggests 
that the correspondence to branched polymers can
be extended to the crumpled phase as well, when
taken with some precaution \cite{bbpt}. Indeed,
the distribution of vertex orders on the minbu
trees has a peak departing from the rest of the
distribution when the number of minbus grows.
Compare the figure \ref{rbb2} and the figure \ref{4d3}. 

The application of the balls-in-boxes model can be extended
beyond the effective theory for minbu trees. Namely,
the model gives also a plausible explanation of the appearance of 
the singular vertices on the simplicial manifold.
If one repeats the same line of arguments 
as in the section on random surfaces, 
one obtains a constrained mean field approximation for
the distribution of the vertex orders of simplicial
manifolds. The partition function for the ensemble 
with $N_0$ vertices and $N_4$ 4-simplices is approximated 
as follows~:
\beq
z(N_4,N_0) = \sum_{\{ q_i\}} p(q_1)\dots p(q_{N_0}) 
\delta ( q_1 + \dots q_{N_0} - 5 N_4)
\eeq
and one ends up with the balls-in-boxes model with
the density $\rho = 5N_4/N_0$. If one lowers  $N_0$,
keeping $N_4$ fixed, the density decreases and one
triggers the transition to the collapsed phase with
singular vertices. The mean field approximation assumes
the independence of orders of vertices as a first 
approximation. The approximation does not give any 
particular form of $p(q)$. Some numerical estimates for the weights 
$p(q)$ were given in \cite{ckr2}. 
In the standard numerical setup one uses the $(\kappa_0,N_4)$
ensemble. In this ensemble, the average $N_0$ grows
with $\kappa_0$. Thus one expects the low
density phase for large $\kappa_0$ and the high density for small
$\kappa_0$. Indeed this is the case. Moreover the $(\kappa_0,N_4)$
ensemble corresponds to the $(\kappa,N)$ ensemble 
of the balls-in-boxes model (\ref{bbccan})
 which has a discontinuous phase transition with 
a double peak distribution of $M/N$ (figure \ref{bb3}).
As we have seen, this is what one observes in 
the numerical data for simplicial gravity, too (figure \ref{4d5}).
Introducing some next order corrections 
to the mean-field approximation by
taking into account the geometrical structure 
of four dimensional simplicial manifolds 
one can explain appearance of the singular 
links as well \cite{bbpt}.

The phase structure of the model discussed in the
present section resembles
the phase structure of random surfaces 
above $D=c=1$. 
The respective line in the $(\alpha,D)$ plane (see figure \ref{2d1})
crosses only
the branched polymer phase and the collapsed phase, exactly
as in four dimensions. We know that in two dimensions, the system
enters the branched polymer phase when the entropy of spiky conformal 
configurations becomes dominant.
The entropy is confronted with the measure term whose dominance
would mean that the system is in the collapsed phase. There is
no other possibility on the line above $D=1$. 
To open the physical window for the gravity phase in two dimensions, 
one has to change the matter content of the theory by decreasing 
the central charge $c=D$ below one. 
This is an important lesson.

One can similarly expect that the phase structure
of four dimensional simplicial gravity 
depends on the matter dressing of the theory \cite{jk,bbkptt}.
Outside the physical window the system is realized 
either by the branched polymer phase or by the collapsed phase.
The question is now how to find the window for the 
gravitational phase. To start with, one can formulate
a more moderate goal, and ask how to prevent the system
from entering the collapsed or the branched polymer 
phase. This problem has been addressed recently \cite{amm,bbkptt}. 
Again it is useful to refer to the analogy with the two dimensional case.
The instability of the Liouville phase is caused by the entropy
of spiky configurations of the conformal field \cite{c,k3,d5}. 
One can now try to repeat the arguments to the effective
action for the conformal mode in four dimensions \cite{amm}. 
The coefficient standing in front of the conformal anomaly~:
\beq
Q^2 = \frac{1}{180} (N_S + \frac{11}{2} N_{F} + 62 N_V -28) \, + Q^2_{grav}
\label{qq}
\eeq
depends on the matter content of the theory through
the number of scalar fields $N_S$, vector fields $N_V$ and fermions 
$N_F$, coupled to gravity  \cite{cd}. 
The $-28$ comes from the ghost sector.
The contribution from gravitons $Q^2_{grav}$ 
has not been calculated since it strongly depends on 
the ultraviolet physics for which the perturbative
treatment fails. The coefficient $Q^2$ plays the
role of the effective central charge in the theory
and it enters the estimates of the free energy of spiky
conformal configurations. The reasoning is analogous
as for the Liouville action in two dimensions \cite{c,k3,d5}.
It turns out, that for $Q^2$ less than a certain
critical value~: $Q^2 < Q^2_{crit}$ the system is dominated 
by the entropy of spikes which means that the system is in the
branched polymer phase\footnote{Note, that in two dimensions the 
analogous inequality is in the opposite direction $c>1$.}. 
Hence, contrary to two dimensions, one expects that the 
fewer degrees of freedom coupled to four dimensional
gravity, the higher is the probability for the system to be 
in the branched polymer phase. 
An inspection of the formula (\ref{qq}) shows that 
the strongest increase of the conformal charge $Q^2$
comes from vector fields. In the paper \cite{amm},
the value of $Q^2_{crit} - Q^2_{grav}$ was estimated
to be of order unity. If this is true, this means that 
by adding few generations of vector
fields, one should be able to prevent the branching  
induced by the entropical instability of 
the conformal factor. Indeed, contrary to the previous 
investigations where some other matter fields were used 
\cite{abjk2,abbjp,ckr3,ckr4}, which did not affect 
significantly the phase structure of simplicial gravity, 
the recent simulations with a varying number
of vector fields \cite{bbkptt} have shown that, when 
three vector fields are minimally coupled to gravity, the 
model has no branched polymer phase. 
Instead a new phase 
is created, where the value of the susceptibility exponent
is negative, similarly as in the Liouville phase of 
the two dimensional gravity. Now we can readdress
the question about the existence of a critical point
in the extended model with the matter fields and 
about the order of the phase transition.

To summarize, statistical physics of four dimensional complexes 
with the simplest geometrical action depending on the number of
points and the number of four simplices is well understood. It is,
however, not related to the continuum physics and we understand 
why. We hope that the problem can be cured by extending the
phase structure of the model by 
adding the appropriate matter fields.

\section*{Summary}

We have reviewed statistical models of random lattices
used as a regularization of the problem of summing over
the internal random geometry of one, two and four dimensional
objects. The degree of difficulty in solving the problem 
grows with the dimensionality of the system, as one
might have naively expected. On the other hand, 
we have shown that there are some common mechanisms 
and features, like the geometrical collapse 
or the existence of the
branched polymer phase, which are almost independent
of the dimensionality of the problem.
Indeed, although the collapsed phase looks slightly
different in one, two or four dimensions,
the primary feature, namely that it is related to 
the existence of singular vertices 
created as a surplus anomaly, is common for 
all cases. The same universality can be found in the
branched polymer phase.

The most interesting part of the theory, {\em ie} 
the physical window, where the discrete models can
be related to the continuum physics, is dimension dependent. 

In two dimensions, in the Liouville phase, related to
the continuum physics, the universal properties
of the model, like the scaling dimensions, are entirely
determined by a single parameter being the conformal 
charge of the theory. There is a discrete series 
of unitary models of conformal matter coupled 
minimally to gravity with the central charge in 
the range between zero and one which have a realization
as statistical models on dynamical triangulations.
The Liouville phase is basically a theory of the conformal
factor.

The situation is more complicated in four dimensions 
where the theory is much more complex and requires
extending the analysis beyond the conformal 
sector. 
But already the analysis of the conformal sector imposes
some restrictions on the combinations of the 
numbers of various fields needed to avoid the 
conformal instability. One expects that there are
some other conditions one has to impose on the number 
of generations of various fields, which lead to 
a specific mixture of fields for which the theory 
is well defined. 

\section*{Acknowledgments}
I am grateful to J. Ambj\o{}rn, P. Bialas, S. Bilke, D. Johnston, 
J. Jurkiewicz, A. Krzywicki, B. Petersson, 
J. Tabaczek and G. Thorleifsson for 
many stimulating discussions and the fruitful collaboration.
A part of my research in this area
was done when I was a fellow of the  Alexander von Humboldt Foundation
and then a DFG fellow at the University of Bielefeld 
and when I was visiting the Niels Bohr Institute in Copenhagen and 
the LPTHE in Orsay. I would like to thank for hospitality in all
these places. The work was partially supported by 
the KBN 2P03B04412 grant.

\end{document}